

%
%
%

\ifx\order\undefined

\catcode`@=11


\newif\if@xrf\@xrffalse   
\def\l@bel #1 #2 #3>>{\expandafter\gdef\csname @@#1#2\endcsname{#3}}
\immediate\newread\xrffile
\def\xrf@n#1#2{\expandafter\expandafter\expandafter
\csname immediate\endcsname\csname #1\endcsname\xrffile#2}
\def\xrf@@n{\if@xrf\relax\else%
  \expandafter\xrf@n{openin}{ = \jobname.xrf}\relax%
  \ifeof\xrffile%
    \message{ no file \jobname.xrf - run again for correct forward
references }%
  \else%
    \expandafter\xrf@n{closein}{}\relax%
    \catcode`@=11 \input\jobname.xrf \catcode`@=12%
  \fi\global\@xrftrue%
  \expandafter\expandafter\csname immediate\endcsname%
  \csname  newwrite\endcsname\xrffile%
  \expandafter\xrf@n{openout}{ = \jobname.xrf}\relax\fi}


\newcount\t@g

\def\order#1{%
  \expandafter\expandafter\csname newcount\endcsname
  \csname t@ghd#1\endcsname\csname t@ghd#1\endcsname=0

  \expandafter\def\csname #1\endcsname##1{\xrf@@n\csname
n@#1\endcsname##1:>}

  \expandafter\def\csname n@#1\endcsname##1:##2>%
    {\def\n@xt{##1}\ifx\n@xt\empty%
     \expandafter\csname n@@#1\endcsname##1:##2:>
     \else\def\n@xt{##2}\ifx\n@xt\empty%
     \expandafter\csname n@@#1\endcsname\unp@ck##1 >:##2:>\else%
     \expandafter\csname n@@#1\endcsname\unp@ck##1 >:##2>\fi\fi}

  \expandafter\def\csname n@@#1\endcsname##1:##2:>%
    {\edef\t@g{\csname t@g#1\endcsname}\edef\t@@ghd{\csname
t@ghd#1\endcsname}%
     \ifnum\t@@ghd=\t@ghd\else\global\t@@ghd=\number\t@ghd\global\t@g=0\fi%
     \ifunc@lled{@#1}{##1}\global\advance\t@g by 1%
       {\def\n@xt{##1}\ifx\n@xt\empty%
       \else\writ@new{#1}{##1}{\pret@g\t@ghead\number\t@g}\expandafter%
       \xdef\csname @#1##1\endcsname{\pret@g\t@ghead\number\t@g}\fi}%
       {\pret@g\t@ghead\number\t@g}%
     \else\def\n@xt{##1}%
       \w@rnmess#1,\n@xt>\csname @#1##1\endcsname%
     \fi##2}\ord@r{#1}}

\def\ord@r#1{%
  \expandafter\expandafter\csname newcount\endcsname
  \csname t@g#1\endcsname\csname t@g#1\endcsname=0

  \expandafter\def\csname ref#1\endcsname##1{%
     \expandafter\each@rg\csname #1c@te\endcsname{##1}}

  \expandafter\def\csname #1c@te\endcsname##1:##2,%
    {\def\n@xt{##2}\ifx\n@xt\empty%
     \csname #1cit@\endcsname##1:##2:,\else%
       \csname #1cit@\endcsname##1:##2,\fi}

  \expandafter\def\csname #1cit@\endcsname##1:##2:,%
    {\def\n@xt{\unp@ck##1 >}\ifunc@lled{@#1}{\n@xt}%
      {\expandafter\ifx\csname @@#1\n@xt\endcsname\relax%
       \und@fmess#1,\n@xt>>>\n@xt<<%
       \else\csname @@#1\n@xt\endcsname##2\fi}%
     \else\csname @#1\n@xt\endcsname##2%
     \fi}}


\def\sporder#1{%

  \expandafter\def\csname #1\endcsname##1{\xrf@@n\csname
n@#1\endcsname##1:>}

  \expandafter\def\csname n@#1\endcsname##1:##2>%
    {\def\n@xt{##1}\ifx\n@xt\empty%
     \expandafter\csname n@@#1\endcsname##1:##2:>
     \else\def\n@xt{##2}\ifx\n@xt\empty%
     \expandafter\csname n@@#1\endcsname\unp@ck##1 >:##2:>\else%
     \expandafter\csname n@@#1\endcsname\unp@ck##1 >:##2>\fi\fi}

  \expandafter\def\csname n@@#1\endcsname##1:##2:>%
    {\edef\t@g{\csname t@g#1\endcsname}%
     \ifunc@lled{@#1}{##1}\global\advance\t@g by 1%
       {\def\n@xt{##1}\ifx\n@xt\empty%
       \else\writ@new{#1}{##1}{\number\t@g}\expandafter%
       \xdef\csname @#1##1\endcsname{\number\t@g}\fi}{\number\t@g}%
     \else\def\n@xt{##1}\w@rnmess#1,\n@xt>\csname @#1##1\endcsname%
     \fi##2}\ord@r{#1}}


\def\each@rg#1#2{{\let\thecsname=#1\expandafter\first@rg#2,\end,}}
\def\first@rg#1,{\callr@nge{#1}\apply@rg}
\def\apply@rg#1,{\ifx\end#1\let\n@xt=\relax%
\else,\callr@nge{#1}\let\n@xt=\apply@rg\fi\n@xt}

\def\callr@nge#1{\calldor@nge#1-\end-}
\def\callr@ngeat#1\end-{#1}
\def\calldor@nge#1-#2-{\ifx\end#2\thecsname#1:,%
  \else\thecsname#1:,\hbox{\rm--}\thecsname#2:,\callr@ngeat\fi}


\def\unp@ck#1 #2>{\unp@@k#1@> @>>}
\def\unp@@k#1 #2>>{\ifx#2@\@np@@k#1\else\@np@@k#1@> \unp@@k#2>>\fi}
\def\@np@@k#1#2#3>{\ifx#2@\@@np@@k#1>\else\@@np@@k#1>\@np@@k#2#3>\fi}
\def\@@np@@k#1>{\ifcat#1\alpha\expandafter\@@np@@@k\string#1>\else#1\fi}
\def\@@np@@@k#1#2>{@#2}


\def\writ@new#1#2#3{\xrf@@n\immediate\write\xrffile
  {\noexpand\l@bel #1 #2 {#3}>>}}


\def\ifunc@lled#1#2{\expandafter\ifx\csname #1#2\endcsname\relax}
\def\und@fmess#1#2,#3>{\ifx#1@%
  \message{ ** error - eqn label >>#3<< undefined - run again ** }\else
  \message{ ** error - #1#2 label >>#3<< undefined - run again ** }\fi}
\def\w@rnmess#1#2,#3>{\ifx#1@%
  \message{ Warning - duplicate eqn label >>#3<< }\else
  \message{ Warning - duplicate #1#2 label >>#3<< }\fi}


\def\t@ghead{}
\newcount\t@ghd\t@ghd=0
\def\taghe@d#1{\gdef\t@ghead{#1}\global\advance\t@ghd by 1}


\order{@qn}


\let\eqno@@=\eqno
\def\eqno(#1){\xrf@@n\eqno@@\hbox{{\rm(}$\@qn{#1}${\rm)}}}

\let\leqno@@=\leqno
\def\leqno(#1){\xrf@@n\leqno@@\hbox{{\rm(}$\@qn{#1}${\rm)}}}

\def\refeq#1{\xrf@@n{{\rm(}$\ref@qn{#1}${\rm)}}}


\def\eqalignno#1{\xrf@@n\displ@y \tabskip=\centering
  \halign to\displaywidth{\hfil$\displaystyle{##}$\tabskip=0pt
   &$\displaystyle{{}##}$\hfil\tabskip=\centering
   &\llap{$\eqaln@##$}\tabskip=0pt\crcr
   #1\crcr}}

\def\leqalignno#1{\xrf@@n\displ@y \tabskip=\centering
  \halign to\displaywidth{\hfil$\displaystyle{##}$\tabskip=0pt
   &$\displaystyle{{}##}$\hfil\tabskip=\centering
    &\kern-\displaywidth\rlap{$\eqaln@##$}\tabskip\displaywidth\crcr
   #1\crcr}}

\def\eqaln@#1#2{\relax\ifcat#1(\expandafter\eqno@\else\fi#1#2}
\def\eqno@(#1){\xrf@@n\hbox{{\rm(}$\@qn{#1}${\rm)}}}


\def\n@@me#1#2>{#2}
\def\numberby#1{\xrf@@n
  \ifx\s@ction\undefined\else
  \expandafter\let\csname\s@@ve\endcsname=\s@ction\fi
  \ifx\subs@ction\undefined\else
  \expandafter\let\csname\subs@@ve\endcsname=\subs@ction\fi
  \numb@rby#1,>#1>}
\def\numb@rby#1,#2>#3>{\def\n@xt{#1}\ifx\n@xt\empty\taghe@d{}\else
  \def\n@xt{#2}\ifx\n@xt\empty\n@by#3>\else\n@@by#3>\fi\fi}
\def\n@by#1>{\ifx\s@cno\undefined\expandafter\expandafter
  \csname newcount\endcsname\csname s@cno\endcsname
  \csname s@cno\endcsname=0\else\s@cno=0\fi
  \xdef\s@@ve{\expandafter\n@@me\string#1>}
  \let\s@ction=#1\def#1{\global\advance\s@cno by 1
  \taghe@d{\number\s@cno.}\s@ction}}
\def\n@@by#1,#2>{\ifx\s@cno\undefined\expandafter\expandafter
  \csname newcount\endcsname\csname s@cno\endcsname
  \csname s@cno\endcsname=0\else\s@cno=0\fi
  \ifx\subs@cst\undefined\expandafter\expandafter
  \csname newcount\endcsname\csname subs@cst\endcsname
  \csname subs@cst\endcsname=0\else\subs@cst=0\fi
  \ifx\subs@cno\undefined\expandafter\expandafter
  \csname newcount\endcsname\csname subs@cno\endcsname
  \csname subs@cno\endcsname=0\else\subs@cno=0\fi
  \xdef\s@@ve{\expandafter\n@@me\string#1>}
  \let\s@ction=#1\def#1{\global\advance\s@cno by 1
  \global\subs@cno=\subs@cst
  \taghe@d{\number\s@cno.}\s@ction}
  \xdef\subs@@ve{\expandafter\n@@me\string#2>}
  \let\subs@ction=#2\def#2{\global\advance\subs@cno by 1
  \taghe@d{\number\s@cno.\number\subs@cno.}\subs@ction}}


\def\numberfrom#1{\ifx\s@cno\undefined\else\n@mberfrom#1,>\fi}
\def\n@mberfrom#1,#2>{\def\n@xt{#2}%
  \ifx\n@xt\empty\n@@f#1>\else\n@@@f#1,#2>\fi}
\def\n@@f#1>{\s@cno=#1\advance\s@cno by -1}
\def\n@@@f#1,#2,>{\s@cno=#1\advance\s@cno by -1%
  \subs@cst=#2\advance\subs@cst by -1}


\def\pret@g{}
\def\prefixby#1{\gdef\pret@g{#1}}



\newcount\r@fcount\r@fcount=0
\newcount\r@fcurr
\newcount\r@fmin
\newcount\r@fmax
\newcount\r@fone
\newcount\r@ftwo
\newif\ifc@te\c@tefalse
\newif\ifr@feat

\def\refto#1{{\rm[}\def\s@p{}\r@fmin=\r@fcount\r@fmax=0%
   \refn@te#1>>\r@fcurr=\r@fmin\advance\r@fcurr by-1\refc@te#1>>{\rm]}}

\def\refn@te#1>>{\refn@@te#1,>>}

\def\refn@@te#1,#2>>{\r@fnote{\expandafter\unp@ck\str@pbl#1 >> >}%
   \def\n@xt{#2}\ifx\n@xt\empty\else\refn@@te#2>>\fi}

\def\r@fnote#1%
  {\ifunc@lled{r@f}{#1}\global\advance\r@fcount by 1\r@fmax=\r@fcount%
   \expandafter\xdef\csname r@f#1\endcsname{\number\r@fcount}%
   \expandafter\gdef\csname r@ftext\number\r@fcount\endcsname%
   {\message{ Reference #1 to be supplied }%
   Reference $#1$ to be supplied\par}%
   \else\expandafter\r@fcurr=\csname r@f#1\endcsname%
   \ifnum\r@fmin<\r@fcurr\else\r@fmin=\r@fcurr\advance\r@fmin by-1\fi%
   \ifnum\r@fmax<\r@fcurr\r@fmax=\r@fcurr\fi\fi}

\def\str@pbl#1 #2>>{#1#2}

\def\refc@te#1>>{\r@featfalse\def\s@ve{}%
  {\loop\ifnum\r@fcurr<\r@fmax\advance\r@fcurr by 1\c@tefalse%
   \expandafter\refc@@te\number\r@fcurr>>#1,>>%
   \ifc@te\expandafter\refe@t\number\r@fcurr>>\fi\repeat\s@ve}}

\def\refc@@te#1>>#2,#3>>{\def\n@xt{\expandafter\unp@ck\str@pbl#2 >> >}%
   \expandafter\refc@@@te\csname r@f\n@xt\endcsname>>#1>>%
   \def\n@xt{#3}\ifx\n@xt\empty\else\refc@@te#1>>#3>>\fi}

\def\refc@@@te#1>>#2>>{\ifnum#2=#1\relax\c@tetrue\fi}

\def\refe@t#1>>{\ifr@feat\ifnum\r@fone=\r@ftwo\res@cond#1>>%
   \else\reth@rd#1>>\fi\else\r@feattrue\ref@rst#1>>\fi}

\def\ref@rst#1>>{\r@feattrue\r@fone=#1\r@ftwo=#1%
   \s@p\expandafter\relax\number\r@fone}%

\def\res@cond#1>>{\advance\r@ftwo by 1\def\n@xt{#1}%
   \expandafter\ifnum\n@xt=\number\r@ftwo%
   \edef\s@ve{,\expandafter\relax\number\r@ftwo}\else,\ref@rst#1>>\fi}%

\def\reth@rd#1>>{\advance\r@ftwo by 1\def\n@xt{#1}%
   \expandafter\ifnum\n@xt=\number\r@ftwo%
   \edef\s@ve{--\expandafter\relax\number\r@ftwo}\def\s@p{,}\else%
   \def\s@p{,}\s@ve\def\s@ve{}\ref@rst#1>>\fi}%


\def\refis#1 #2\par{\def\n@xt{\unp@ck#1 >}\r@fis\n@xt>>#2>>}
\def\r@fis#1>>#2>>{\ifunc@lled{r@f}{#1}\else
   \expandafter\gdef\csname r@ftext\csname
r@f#1\endcsname\endcsname{#2\par}\fi}


\def\listreferences{\global\r@fcurr=0%
  {\loop\ifnum\r@fcurr<\r@fcount\global\advance\r@fcurr by 1%
   \numr@f\number\r@fcurr>>\csname r@ftext\number\r@fcurr\endcsname>>%
   \repeat}}

\def\numr@f#1>>#2>>{\vbox{\noindent\hang\hangindent=30truept%
   {\hbox to 30truept{\rm[#1]\hfill}}#2}\smallskip\par}


\def\printlabels{\global\@xrftrue\def\s@me##1{$##1$}
  \def\@qn##1{##1}\def\refeq##1{{\rm(}$##1${\rm)}}\refbylabel
  \def\numberby##1{\relax}\def\numberfrom##1{\relax}
  \def\listreferences{\relax}\def\referencefile{\relax}
  \def\order##1{\expandafter\let\csname ##1\endcsname=\s@me
                \expandafter\let\csname ref##1\endcsname=\s@me}
  \def\sporder##1{\expandafter\let\csname ##1\endcsname=\s@me
                \expandafter\let\csname ref##1\endcsname=\s@me}}

\def\refbylabel{\def\refto##1{[$##1$]}%
  \def\refis##1 ##2\par{\numr@f$##1$>>##2>>}\def\numr@f##1>>##2>>%
  {\noindent\hang\hangindent=30truept{{\rm[##1]}\ }##2\par}}


\def\beginsection#1\par{\vskip0pt plus.3\vsize\penalty-250
  \vskip0pt plus -.3\vsize\bigskip\vskip\parskip
  \leftline{\bf#1}\nobreak\smallskip\noindent}

\catcode`@=12
%
\catcode`@=11
\newif\ifnews@ct\news@ctfalse

\magnification 1200
\hsize=15.1truecm
\vsize=23truecm
\hoffset=0.5truecm
\voffset=1truecm
\parskip=2pt
\nopagenumbers
\pageno=0
\headline={\ifnum\pageno=0{}\else \hdline\fi}
\def\hdline{\ifodd\pageno\rightheadline \else\leftheadline\fi}
\def\rightheadline{\tenrm\hfil{\it \rhead}\hfil\folio}
\def\leftheadline{\tenrm\folio\hfil{\it \lhead}\hfil}
\newcount\secno\secno=0
\def\section#1#2\par#3\par
   {\vskip0pt plus.1\vsize\penalty-250
    \vskip0pt plus-.1\vsize\bigskip
    \if#1*\noindent{\bf#2}\else
    \advance\secno by 1\subsecno=0\news@cttrue
    \noindent\hbox to \parindent{\bf\number\secno.\hfil}{\bf#1}\fi
    \par\vskip-\parskip\medskip#3\news@ctfalse\par}
\newcount\subsecno\subsecno=0
\def\subsection#1
   {\advance\subsecno by 1
   \ifnews@ct\else\smallskip\fi
   \noindent\number\secno.\number\subsecno\hskip8pt{\sl #1}\quad}

%
%

\font\tenmib=cmmib10
\font\sevenmib=cmmib10 scaled 700
\font\fivemib=cmmib10 scaled 500
\newfam\mibfam
\textfont\mibfam=\tenmib
\scriptfont\mibfam=\sevenmib
\scriptscriptfont\mibfam=\fivemib
\def\mib{\fam\mibfam\tenmib}

%
%
\newif\ifpr@@f\pr@@ffalse
\def\proof{\pr@@ftrue\smallskip\noindent {\sl Proof.}\quad}
\def\indproof{\pr@@ftrue\smallskip\noindent {\sl Proof.} (by
induction)\quad}
\def\qed{\pr@@ffalse\ifmmode\qquad{\hbox{\Fsquare(.2cm,{})}}
\else\qquad$\Fsquare(.2cm,{})$\par\noindent\fi}
\def\Fsquare(#1,#2)%
{\hbox{\vrule$\hskip-0.4pt\vcenter to #1{\normalbaselines\m@th%
  \hrule\vfil\hbox to #1{\hfill$#2$\hfill}\vfil\hrule}$\hskip-0.4pt\vrule}}
  \def\m@th{\mathsurround=0pt}

\def\point(#1){\ifpr@@f\pr@@ffalse\else\par\fi\noindent%
  \hbox to \parindent{\rm(#1)\hfill}\ignorespaces}

%
\sporder{defn}
\let\def@=\defn\let\refdef@=\refdefn
\def\defn#1#2.#3\enddef%
  {\csname proclaim\endcsname Definition
   \def@{#1}#2.#3\par\vskip-\parskip\noindent}
\def\refdef#1{Definition \refdef@{#1}}

\sporder{thm}
\let\thm@=\thm\let\refthm@=\refthm
\def\thm#1#2.#3\endthm%
  {\csname proclaim\endcsname Theorem
   \thm@{#1}#2.#3\par\vskip-\parskip\noindent}
\def\refthm#1{Theorem \refthm@{#1}}

\sporder{prop}
\let\prop@=\prop\let\refprop@=\refprop
\def\prop#1#2.#3\endprop%
  {\csname proclaim\endcsname Proposition
   \prop@{#1}#2.#3\par\vskip-\parskip\noindent}
\def\refprop#1{Proposition \refprop@{#1}}

\order{lem}
\let\lem@=\lem\let\reflem@=\reflem
\def\lem#1#2.#3\endlem%
  {\csname proclaim\endcsname Lemma
   \lem@{#1}#2.#3\par\vskip-\parskip\noindent}
\def\reflem#1{Lemma \reflem@{#1}}
\def\reflems#1#2{Lemmas \reflem@{#1}\hskip.2em\&\hskip.1em\reflem@{#2}}

\order{cor}
\let\cor@=\cor\let\refcor@=\refcor
\def\cor#1#2.#3\endcor%
  {\csname proclaim\endcsname Corollary
   \cor@{#1}#2.#3\par\vskip-\parskip\noindent}
\def\refcor#1{corollary \refcor@{#1}}
\def\refCor#1{Corollary \refcor@{#1}}
%
%
\def\etal{{\it et.al.}}
\def\ie{{\it i.e.}}
\def\id{{\rm id\/}}

\def\mod{\mathop{\rm mod\/}}
\def\End{\mathop{\rm End\/}}
\def\Hom{\mathop{\rm Hom\/}}
\def\Ker{\mathop{\rm Ker\/}}
\def\et{\tilde{e}}
\def\ft{\tilde{f}}

\def\ap{{S}}
\def\st{{\sf t}}
\def\ce{\gamma}

\def\mapright#1{\mathop{\longrightarrow}\limits^{#1}}
\def\HX#1{\def\next{#1}\ifx\next\empty H_{\rm XXZ}
  \else H_{{\rm XXZ},#1}\fi}
\def\HC#1{\def\next{#1}\ifx\next\empty H_{\rm CTM}
  \else H_{{\rm CTM},#1}\fi}
\def\ba{{\mib a}}
\def\bl{{\mib l}}
\def\slt{\gs\gl_2}
\def\slth{\widehat{\slt}}
\def\U{U_q\bigl(\slt\bigr)}
\def\Up{U'_q\bigl(\slth\bigr)}
\def\Uq{U_q\bigl(\slth\bigr)}
\ifx\Bbb\undefined
  \def\BC{{\bf C}}
  \def\BL{{\bf L}}
  
  \def\BQ{{\bf Q}}
  \def\BR{{\bf R}}
  \def\BZ{{\bf Z}}
\else
  \def\BC{{\Bbb C}}
  \def\BL{{\Bbb L}}
  
  \def\BQ{{\Bbb Q}}
  \def\BR{{\Bbb R}}
  \def\BZ{{\Bbb Z}}
\fi
\ifx\goth\undefined
  \def\gs{{\sl s}}
  \def\gl{{\sl l}}
  \def\gH{{\sl H}}
\else
  \def\gs{{\goth s}}
  \def\gl{{\goth l}}
  \def\gH{{\goth H}}
\fi
\def\CX{\BC^{\times}}
%
\def\figure#1#2{\vglue 0.4cm plus 0.1cm
\centerline{\epsfbox{#1}}\centerline{#2}
\vglue 0.6cm plus 0.1cm minus 0.1cm}
%
%
%
  \font\tensans=cmss10 scaled 1050
  \font\sevensans=cmss10 scaled 735
  \font\fivesans=cmss10 scaled 525
  \newfam\sansfam
  \textfont\sansfam=\tensans
  \scriptfont\sansfam=\sevensans
  \scriptscriptfont\sansfam=\fivesans
  \def\sf{\fam\sansfam\tensans}
  \def\mbXXZ{{\mib X\kern-2pt X\kern-2pt Z}}

\catcode`@=12
%
\catcode`@=11
\newif\ifs@p
\def\refjl#1#2#3#4%
  {#1\def\l@st{#1}\ifx\l@st\empty\s@pfalse\else\s@ptrue\fi%
   \def\l@st{#2}\ifx\l@st\empty\else%
   \ifs@p, \fi{\frenchspacing\sl#2}\s@ptrue\fi%
   \def\l@st{#3}\ifx\l@st\empty\else\ifs@p, \fi{\bf#3}\s@ptrue\fi%
   \def\l@st{#4}\ifx\l@st\empty\else\ifs@p, \fi#4\s@ptrue\fi%
   \ifs@p.\fi\hfill\penalty-9000}
\def\refbk#1#2#3%
  {#1\def\l@st{#1}\ifx\l@st\empty\s@pfalse\else\s@ptrue\fi%
   \def\l@st{#2}\ifx\l@st\empty\else%
   \ifs@p, \fi{\frenchspacing\sl#2}\s@ptrue\fi%
   \def\l@st{#3}\ifx\l@st\empty\else\ifs@p, \fi#3\s@ptrue\fi%
   \ifs@p.\fi\hfill\penalty-9000}
\catcode`@=12
%
%

\def\CMP{Comm. Math. Phys.}

\def\IJMPA{Int. J. Mod. Phys. A}

\def\IM{Inv. Math.}

\def\JPA{J. Phys. A: Math. Gen.}

\def\JSP{J. Stat. Phys.}
\def\LMP{Lett. Math. Phys.}
\def\MA{Math. Ann.}

\def\NPB{Nucl. Phys. B}

\def\PLA{Phys. Lett. A}

\def\RIMS{RIMS preprint}

  \input epsf.tex
\fi
%
%
\def\ltitle{Quantum Loop Modules and Quantum Spin Chains}

\def\lauthor{Daniel Altschuler and Brian Davies}

\def\laddress{Laboratoire de Physique Th\'eorique ENSLAPP
${}^*$\cr
Ecole Normale Sup\'erieure de Lyon,\cr
46, all\'ee d'Italie,\cr
69364 Lyon Cedex 07, France \cr
{}\cr
and\cr
{}\cr
Department of Mathematics, the Faculties,\cr
Australian National University, \cr
GPO Box 4, Canberra, ACT 2601, Australia \cr}

\def\labstract{
We construct level-0 modules of the quantum affine algebra $\Uq$,
as the $q$-deformed version of the Lie algebra loop module
construction.
We give necessary and sufficient conditions for the modules to be
irreducible.
We construct the crystal base for some of these modules and find
significant differences from the case of highest weight modules.
We also consider the role of loop modules in the recent scheme for
diagonalising certain quantum spin chains using their $\Uq$ symmetry.}

%
%
\def\lhead{\lauthor}
\def\rhead{\ltitle}
\vglue 1cm
\ifx\twelvebf\undefined\font\twelvebf=cmbx10 scaled 1200 \fi
\centerline{\twelvebf
\vbox{\halign{\hfil # \hfil\cr\ltitle\crcr}}}
\vglue 1cm
\ifx\twelveit\undefined\font\twelveit=cmti10 scaled 1200 \fi
\centerline{\twelveit
\vbox{\halign{\hfil # \hfil\cr\lauthor\crcr}}}
\vglue 1cm
\centerline{\vbox{
\halign{\hfil # \hfil\cr
\laddress\crcr}}}
\vglue 2cm
{\narrower{\noindent\bf Abstract.}\quad\labstract\par}
\vglue 1cm
\centerline{In memory of Rolf Adams.}
\vglue 1cm
\hbox{~}\hfill MRR2/93\break
\hbox{~}\hfill ENSLAPP-L-419/93\break
\hbox{~}\hfill May 1993\break
\footnote{}
{${}^*$~URA 1436 du CNRS, associ\'ee \`a l'Ecole Normale
Sup\'erieure de Lyon et au laboratoire d'Annecy-le-Vieux
de Physique des Particules.}
\vfill \eject
%
%
\numberby{\section}

\section{Introduction}

Since the advent of quantum mechanics, representation theory and
mathematical physics have enjoyed close ties --- indeed, in some areas
this is so self-evident as to go almost un-noticed.
In the subject of exactly solved models of two-dimensional statistical
mechanics, it has not always been so.
But developments in the last decade have been rapid.
The quantum inverse scattering method  --- an algebraic
implementation of the Bethe Ansatz and associated Yang-Baxter
equations --- culminated in the theory of quantum groups
\refto{J85,Dr87,Dr88,J89,J92}.
Representation theory of infinite dimensional
Lie algebras is paramount in Conformal Field Theory
\refto{BPZ84,FC91}.
In the massive case, Corner transfer matrices (CTMs), invented by
Baxter \refto{Bax76,Bax77}, pointed to the involvement once more of
representation theory, through the appearance of character formulae
\refto{DJKMNO87,DJKMNO89}.

In \refto{DFJMN92} was given a new scheme for diagonalising the
XXZ (six-vertex) Hamiltonian spin chain, in the anti-ferromagnetic
regime, using the Quantum affine algebra $\Uq$.
The aproach of that paper has been extended to higher spin chains
\refto{IIJMNT92}, to the higher rank case \refto{DO93}
and to the SOS models of
Andrews, Baxter and Forrester \refto{ABF84,JMO92}.
In the six-vertex case, the scheme has been used to give general
expressions for the $n$-point correlation functions
\refto{JMMN92}: it also provides a means to derive $q$-difference
equations for correlation functions for all the models
\refto{JMN92,FJMMN93}.
The six-vertex model, and its higher-spin and higher-rank
generalisations, are related to the quantum affine algebra $\Up$ of
Drinfeld and Jimbo \refto{Dr87,J89}, since their Boltzmann weights
form the $R$-matrices which intertwine tensor products of
finite-dimensional representations.
For the SOS models the relation is more technical: the Boltzmann
weights are the connection matrices which intertwine vertex operators
--- themselves intertwiners --- discovered by Frenkel and Reshetikhin
\refto{FR92}.
An important ingredient in the use of the quantum
affine algebra for the solution of these models is the fact that the
eigenvectors of their CTMs may be idendified with the weight vectors
of level-$k$ infinite-dimensional modules, and their CTMs
with the derivation operator $d$ \refto{FM92,Dav93}.
In this way, physical calculations are reduced to problems of
representation theory.

CTMs act on a semi-infinite spin chain rather than the full infinite
chain of the row transfer matrix (RTM) and its associated Hamiltonian.
The scheme presented in \refto{DFJMN92} is that one selects an
arbitrary point in the infinite chain, and then employs the
eigenvectors of a level-1 module and its dual (which is
level $-1$) to represent the states of the left and right hand
semi-infinite parts, respectively.
The state space of the entire chain is then a tensor product
and is a level-0 module.
This tensor product is highly reducible; the most obvious
reduction being the decompositions into $n$-particle states.
One may shift the selected point at which translational symmetry
is broken, one site at a time, using the theory of quantum vertex
operators \refto{FR92}.
This gives a viable representation-theoretic realisation of the
translation operator $T$.
Since the derivation $d$ was already identified with a Hamiltonian
spin chain (the CTM) whose coefficients are linear in the
position along the chain, the usual Hamiltonian spin chain becomes
identified with a multiple of the operator $(TdT^{-1}-d)$.
We shall give some further relevant details, for $\HX{}$, in a
later section of this paper, but otherwise we refer the reader to
the cited references.

These developments highlight the importance of further study of
level-0 \break$\Uq$ modules.
In the Lie algebra case it is shown in \refto{Ch86,CP86} that the
only integrable level-0 modules, with finite dimensional weight
spaces, are loop modules (up to isomorphism).
Since our interest is in the generic case of $\Uq$, we confine our
attention here to loop modules.
They are, of course, constructed by defining an appropriate module
action on the affinization of a finite dimensional $\U$ module ---
see \refeq{3.1} below.
For the most part, this paper considers the fundamental
properties of the loop modules, such as their generation
from ``highest weight'' components, reducibility, tensor
products and the construction of a crystal base.
We do, however, return to the problem which motivated this study
in section 6 of the paper.
There we apply our results to a better understanding of the
diagonalisation scheme for $\HX{}$, referred to above.

The plan of the paper is as follows.
In section 2 we give the necessary definitions of the quantum
algebras and the presentations employed herein.
Section 3 defines loop modules and their character function: we also
derive an important determinant formula to be used later.
Section 4 is devoted to questions about the fundamental
structure of the loop modules and their tensor products.
In section 5 we construct a crystal base for a subset of the
loop modules. We show that the loop modules do not enjoy the nice
existence and uniqueness properties
(modulo the crystal lattice) of a crystal base, which hold
for the highest weight modules.
Section 6 returns to the problem which motivated this work: the
diagonalisation of $\HX{}$ using the quantum affine symmetry.
We give some new results, particularly about the preservation of the
crystal lattice by the particle creation operators.
Some concluding comments are made in section 7.
\section{Notations}

\subsection{Definition of $\Uq$.}
We follow Jimbo \refto{J92}.
$\U$ is an associative algebra generated by $e$, $f$, $t$, which
satisfy the defining relations
$$
t e t^{-1} = q^2 e, \qquad
t f t^{-1} = q^{-2} f, \qquad
[e,f]={t-t^{-1}\over q-q^{-1}}.
\eqno(2.0)
$$
The representations we use are built on the basic
$(l+1)$-dimensional irreducible $\U$ modules $V_l$,
with weight vectors $v^l_k$
$(k=0,\ldots,l)$, defined by the module action
$$
\eqalign{
e\cdot v^l_k &= [l-k+1] v^l_{k-1}, \cr
f\cdot v^l_k &= [k+1] v^l_{k+1}, \cr
t\cdot v^l_k &= q^{l-2k} v^l_k, \cr}
\eqno(2.0a)
$$
where $[n] = (q^n-q^{-n})/(q-q^{-1})$, $v^l_0=0$ is the highest
weight vector, and $v^l_k=0$ if $k<0$ or $k>l$.

For $\Up$ the generators are $e_i$, $f_i$, $t_i$,
$(i=0,1)$, and they satisfy
$$
\eqalignno{
&t_i e_j = q^{A_{ij}} e_j t_i, \qquad\qquad
t_i f_j = q^{-A_{ij}} f_j t_i, \cr
&t_i t_j =t_j t_i, \qquad\qquad\qquad
[e_i,f_j]=\delta_{ij}{t_i-t_i^{-1}\over q-q^{-1}}, &(2.1) \cr
e_i^3 e_j - (q^2+&1+q^{-2}) e_i^2 e_j e_i +
(q^2+1+q^{-2}) e_i e_j e_i^2 - e_j e_i^3 = 0,
\hskip 9pt (i\not=j), \cr
f_i^3 f_j - (q^2+&1+q^{-2}) f_i^2 f_j f_i +
(q^2+1+q^{-2}) f_i f_j f_i^2 - f_j f_i^3 = 0,
\hskip 4pt (i\not=j). \cr}
$$
where $A_{ij}$ is the generalised Cartan matrix for the affine Lie
algebra $\slth$
$$
A_{ij}=\pmatrix{2&-2\cr-2&2\cr}.
\eqno(2.2)
$$
Finally, the full quantum affine algebra $\Uq$ is obtained by adding the
generator $q^d$.
$d$ is the derivation, for which
$$
[d,e_i]=\delta_{i,0}e_i, \qquad
[d,f_i]=-\delta_{i,0}f_i, \qquad
[d,t_i]=0.
\eqno(2.5)
$$

Formulae for the crystal base depend on the normalisation of
the roots via the parameter $q_i=q^{(\alpha_i,\alpha_i)}$.
Since we are only interested in the case of $\Uq$, we
follow \refto{DFJMN92} and choose $(\alpha_i,\alpha_i)=1$.
Consequently, our conventions for $\slth$ are as follows.
The Cartan subalgebra $\gH$ is spanned by $\{h_0,h_1,d\}$ and
$\alpha_0$, $\alpha_1$ are the roots.
They are related to the fundamental weights by
$\alpha_0=2\Lambda_0-2\Lambda_1+\delta$,
$\alpha_1=2\Lambda_1-2\Lambda_0$;
we also write $\rho=\Lambda_0+\Lambda_1$.
The invariant form on $\gH^*$ is given by
$(\Lambda_i,\Lambda_j)=\delta_{i1}\delta_{j1}/4$,
$(\Lambda_i,\delta)=1/2$,
$(\delta,\delta)=0$.
The weight lattice and its dual are
$P=\BZ\Lambda_0\oplus\BZ\Lambda_1\oplus\BZ\delta$ and
$P^*=\BZ h_0\oplus\BZ h_1\oplus\BZ d$, with
$\langle\Lambda_i,h_j\rangle=\delta_{ij}$,
$\langle\Lambda_i,d\rangle=0$, $\langle\delta,h_i\rangle=0$,
$\langle\delta,d\rangle=1$.
We identify $P^*$ with a subset of $P$ via $(~,~)$,
so that  $2\alpha_i=h_i$ and $4\rho=4d+h_1$.

\subsection{Drinfeld's realisation.}
We shall require the realisation of $\Up$ due to Drinfeld
\refto{Dr88}.
$\Up$ is the associative algebra over $\BC$
with generators $K^{\pm1}$, $\{x^\pm_k\mid k\in\BZ\}$,
$\{h_k\mid k\in\BZ_{\not=0}\}$, and central elements
$\ce^{\pm1}$, satisfying the following defining relations:
$$
\eqalign{
&[h_k,h_l]=\delta_{k,-l}
{[2k](\ce^k-\ce^{-k})\over k(q-q^{-1})}, \cr
&[K,h_k]=0, \cr
&Kx^\pm_kK^{-1}=q^{\pm2}x^\pm_k, \cr
&[h_k,x^\pm_l]= \pm k^{-1}[2k] \ce^{\mp(k+|k|)/2} x^\pm_{k+l}, \cr
&x^\pm_{k+1}x^\pm_l-q^{\pm2}x^\pm_lx^\pm_{k+1}=
q^{\pm2}x^\pm_kx^\pm_{l+1}-x^\pm_{l+1}x^\pm_k, \cr
&[x^{+}_k,x^{-}_l]=
{\ce^{k-l}\psi_{k+l}-\phi_{k+l}\over
q-q^{-1}}, \cr}
\eqno(2.11)
$$
where only $\{\psi_k\mid k\in\BZ_{\ge0}\}$ and
$\{\phi_k\mid k\in\BZ_{\le0}\}$ are non-zero; they are given by the
formal expansions
$$
\eqalign{
\psi(u)=\sum_{k=0}^\infty \psi_k u^k
&=K\exp\left((q-q^{-1})\sum_{k=1}^\infty h_k u^k\right), \cr
\phi(u)=\sum_{k=0}^\infty \phi_{-k} u^{-k}
&=K^{-1}\exp\left(-(q-q^{-1})\sum_{k=1}^\infty
h_{-k}u^{-k}\right). \cr}
\eqno(2.12)
$$
The isomorphism between the two presentations \refeq{2.1,2.11} is
given by
$$
\eqalign{
&e_0\mapsto x^-_1 K^{-1}, \cr &e_1\mapsto x^+_0, \cr}\qquad
\eqalign{
&f_0\mapsto \ce K x^+_{-1}, \cr &f_1\mapsto x^-_0, \cr}\qquad
\eqalign{
&t_0\mapsto \ce K^{-1}, \cr &t_1\mapsto K. \cr}
\eqno(2.15)
$$


For $\Uq$ we must add the derivation $d$.
The isomorphism \refeq{2.15} gives the commutators
of $d$ with $x^\pm_0$, $x^{+}_{-1}$, $x^{-}_1$ and $K$.
It is readily checked that these imply
$$
\eqalign{
[d,x^\pm_k]&=k x^\pm_k, \quad (k\in\BZ), \cr
[d,h_k]&=k h_k, \,\quad (k\in\BZ_{\not=0}), \cr}\qquad
\eqalign{
[d,\psi_k]&=k\psi_k, \quad (k\in\BZ_{\ge0}), \cr
[d,\phi_k]&=k\phi_k, \quad (k\in\BZ_{\le0}). \cr}
\eqno(2.17)
$$

\subsection{Co-algebra structure.}
We define a coproduct $\Delta$ and an antipode $\ap$ for the
presentation \refeq{2.1} by
$$
\eqalign{
&\Delta(e_i)=e_i\otimes 1+t_i\otimes e_i,\cr
&\Delta(f_i)=f_i\otimes t_i^{-1}+ 1\otimes f_i,\cr
&\Delta(t_i)=t_i\otimes t_i \cr}\qquad
\eqalign{
&\ap(e_i)=-t_i^{-1}e_i, \cr
&\ap(f_i)=-f_it_i,\cr
&\ap(t_i)=t_i^{-1}. \cr}
\eqno(2.6)
$$
$\Delta$ is related to the co-product used in Chari and Pressley
\refto{CP91} by transposition; in the terminology of Kashiwara
\refto{Ka90} it is the upper co-product.
No general formula for the co-product is known in the Drinfeld
presentation.
However, some partial results are given in \refto{CP91}.
We recall them here for convenience.
Denote by $H$ the subalgebra of $U=\Uq$ generated by $d$, $\ce$,
$K$, $h_k$, $(k\in\BZ_{\not=0})$, let $N_\pm$ be the subalgebras
generated by the $x^\pm_k$, $(k\in\BZ)$, and $X_\pm$ the linear
span of the $x^\pm_k$.
Then we have that $U=N_{-}HN_{+}$.
Furthermore, it is shown in \refto{CP91}, proposition 4.4, that the
co-product $\Delta$ satisfies the following:
\point(i)
$\mod(UX_{-}\otimes UX_{+}^2)$,
$$
\eqalign{
\Delta(x^{+}_k)&=
x^{+}_k\otimes1
+\sum_{i=0}^k \psi_i\otimes x^{+}_{k-i}, \quad (k\ge0), \cr
\Delta(x^{+}_{-k})&= x^{+}_{-k}\otimes1 
+\sum_{i=0}^{k-1} \phi_{-i}\otimes x^{+}_{i-k}, \quad(k>0), \cr}
\eqno(2.7)
$$
\point(ii)
$\mod(UX_{-}^2\otimes UX_{+})$,
$$
\eqalign{
\Delta(x^{-}_k)&=
1\otimes x^{-}_k
+\sum_{i=0}^{k-1} x^{-}_{k-i}\otimes\psi_i, \quad(k>0), \cr
\Delta(x^{-}_{-k})&=
1\otimes x^{-}_{-k}
+\sum_{i=0}^k x^{-}_{i-k}\otimes\phi_{-i}, \quad(k\ge0), \cr}
\eqno(2.8)
$$
\point(iii)
$\mod(UX_{-}\otimes UX_{+}+UX_{+}\otimes UX_{-})$,
$$
\eqalign{
\Delta(\psi_k)&=
\sum_{i=0}^k \psi_i\otimes \psi_{k-i}, \quad (k\ge0), \cr
\Delta(\phi_{-k})&=
\sum_{i=0}^k \phi_{-i}\otimes \phi_{i-k}, \quad (k\ge0), \cr}
\eqno(2.9)
$$
\section{Quantum Loop Modules}

\subsection{Definitions.}
Finite dimensional representations of $\Up$ are constructed
by first defining an evaluation map depending on a parameter $a\in\CX$
(\ie\ $a\not=0$) \refto{J86,CP91}.
Let $\BL=\BC[\st,\st^{-1}]$ be the ring of Laurent polynomials in the
formal variable $\st$.
If $V$ is a vector space over $\BC$, we define the loop space
$\BL(V)$ as
$$
\BL(V)=\BL\otimes V
\eqno(3.1)
$$
as in \refto{CP86}.
The evaluation map which we employ herein for the construction of
loop modules is defined, for the presentation \refeq{2.1}, as follows:
\defn{3.1}.
For any $a\in\CX$ let $\varphi_a:\Up\longrightarrow \BL(\U)$ be
given by
$$
\eqalign{
\varphi_a(e_0)&=\st\otimes af, \cr
\varphi_a(e_1)&=1\otimes e, \cr
\varphi_a(t_0)&=1\otimes K^{-1}, \cr}\qquad
\eqalign{
\varphi_a(f_0)&=\st^{-1}\otimes a^{-1}e, \cr
\varphi_a(f_1)&=1\otimes f, \cr
\varphi_a(t_1)&=1\otimes K. \cr}
\eqno(3.2)
$$
\enddef
It is easy to check that:
\lem{3.1}.
$\varphi_a$ is an algebra homomorphism and the image of
the Drinfeld generators $\ce$, $x^\pm_k$, is given by
$$
\eqalign{
\varphi_a(\ce)&=1, \cr
\varphi_a(x^+_k)&=\st^k\otimes a^k K^k e, \cr
\varphi_a(x^-_k)&=\st^k\otimes a^k f K^k.
\qquad\qquad\qquad\qquad\cr}
\eqno(3.3)
$$
\endlem

We use $\varphi_a$ to construct loop representations of $\Uq$.
Since $\varphi_a(\ce)=1$, these modules will be of level 0.
\defn{3.2}.
Let $(\pi,V)$ be a representation of $\U$.
Then we construct the loop representation $\hat{\pi}_a$
of $\Up$, $a\in\CX$,
on the space $\BL(V)$ as follows:
$$
\hat{\pi}_a=(\id\otimes\pi)\circ\varphi_a,
\eqno(3.4)
$$
\ie, $\hat{\pi}_a$ is a composition of maps
$$
\Up\mapright{\varphi_a}\BL(\U)\mapright{\id\otimes\pi}
\BL(\End(V))=\BL\otimes\End(V).
\eqno(3.5)
$$
This representation $\hat{\pi}_a$ extends to a representation
of $\Uq$, by setting:
$$
\hat{\pi}_a(d) = \st {d \over d\st} \otimes 1.
\eqno()
$$
\enddef
We define more general quantum loop modules by a simple
generalisation of the loop group construction given in \refto{CP86}.
\defn{3.3}.
Let $(\pi^{l_i},V_{l_i})$,
$(i=1,\ldots,k)$, be irreducible $(l_i+1)$-dimensional highest weight
representations of $\U$ and let $a_i\in\CX$.
Then the module $V(\bl;\ba)=
\BL(V_{l_1}\otimes\cdots\otimes V_{l_k})$ is defined via the
$\Uq$ action
$$
\pi_{\bl,\ba}=P_{\BL}\circ
(\hat{\pi}_{a_1}\otimes\cdots\otimes\hat{\pi}_{a_k})\circ\Delta_k
\eqno(3.6)
$$
where $\Delta_k$ is the $k$-fold iterated co-product and $P_{\BL}$
is the linear map
$$
P_{\BL}:\BL(\End(V))\otimes\cdots\otimes \BL(\End(V))\longrightarrow
\BL(\End(V)\otimes\cdots\otimes\End(V))
\eqno(3.7)
$$
which identifies the powers of the formal variable $\st$ in each factor.
\enddef

We remark that the action of $d$ on loop modules is given
by $d\cdot(\st^n\otimes v)=n(\st^n\otimes v)$.
One could define twisted loop modules by modifying this action to
$d\cdot(\st^n\otimes v)=(n+c)(\st^n\otimes v)$, $c\in\CX$, but we
shall not choose to make this generalisation.
Also, one can use arbitrary highest weight $\U$ modules, but we shall
not consider this complication either.
Obviously one recovers finite-dimensional representations of
$\Up$ by dropping the derivation and setting $\st=1$.

\subsection{Dual modules.}
We shall need dual loop modules, which we now define.
First, given a representation $(\pi,V)$ of $\U$, the dual
representation $(\pi^*,V^*)$ is defined by the action
$$
\langle x\cdot u^*,v\rangle=\langle u^*,\ap(x)\cdot v\rangle, \qquad
\forall~u^*\in V^*,~u\in V,~x\in\U.
\eqno(3.41)
$$
Recall also that the loop space $\BL(V)=\BL\otimes V$ is
turned into a representation $(\hat{\pi}_a,\BL(V))$ of $\Uq$
by pulling back a representation $(\pi,V)$ of $\U$ using
the evaluation map $\varphi_a$.
We can use the dual space $V^*$ in the same way.
It is easy to check that
$$
(\id\otimes\ap)\circ\varphi_{q^2a}=\varphi_a\circ\ap,
\eqno(3.42)
$$
where on the left hand side the antipode is acting in $\U$
and on the right hand side it acts in $\Up$.
The definition of a dual loop module may now be given:
\defn{3.2a}.
Let $(\pi,V)$ be a representation of $\U$ and $(\pi^*,V^*)$ its dual.
Then we construct the loop representation $\hat{\pi}^*_a$, $a\in\CX$,
on $\BL(V^*)$ as follows:
$$
\hat{\pi}^*_a=(\id\otimes\pi^*\circ S^{-1})\circ\varphi_a\circ\ap
\eqno(3.4a)
$$
\enddef
In constructing the dual to \refdef{3.3}, one should note that taking
the dual of a tensor product reverses the order of the factors and we
want the dual of $\BL(V_{l_1}\otimes\cdots\otimes V_{l_k})$ to act on
$\BL((V_{l_1}\otimes\cdots\otimes V_{l_k})^*)$.
\defn{3.3a}.
Let $(\pi^{l_i},V_{l_i})$,
$(i=1,\ldots,k)$, be $(l_i+1)$-dimensional highest weight
representations of $\U$ and let $a_i\in\CX$.
Then the module $V^*(\bl;\ba)=
\BL(V^*_{l_k}\otimes\cdots\otimes V^*_{l_1})$ is defined via the
$\Uq$ action
$$
\pi^*_{\bl,\ba}=P_{\BL}\circ
(\hat{\pi}^*_{a_k}\otimes\cdots\otimes\hat{\pi}^*_{a_1})\circ\Delta_k
\eqno(3.43)
$$
with $\Delta_k$ and $P_{\BL}$ as before.
\enddef
Observe that $V^*(\bl;\ba)\subset (V(\bl;\ba))^*$.
\refeq{3.42} is an equality between maps
$\Up\longrightarrow\BL\otimes\U$.
It follows that if
$(\bl,\ba)$ and $(\bl^\prime,\ba^\prime)$ are
related to each other by reversal of order,
then there is an isomorphism of loop modules
$$
V^*(\bl,\ba)\mapright{\sim}V(\bl^\prime,q^2\ba^\prime)
\eqno(3.44)
$$

\subsection{Character formula.}
Using the formulae for $\Delta(\psi_m)$,
$\Delta(\phi_{-m})$ and $\Delta(x^+_k)$ given in section 2.3,
it is easy to see that the ``highest weight components''
$\Omega_{\bl,n}=\st^n\otimes u^{l_1}_0\otimes\cdots u^{l_k}_0$
are eigenvectors of $\psi_m$ and $\phi_{-m}$, $(m\in\BZ_{\ge0})$,
and are annihilated by the subalgebra $N_{+}$.
Write
$$
\eqalign{
\psi_m\cdot\Omega_{\bl,n} &=
\chi_{\bl;\ba}(\psi_m)\Omega_{\bl,n}, \cr
\phi_{-m}\cdot\Omega_{\bl,n} &=
\chi_{\bl;\ba}(\phi_{-m})\Omega_{\bl,n}, \cr
N_{+}\cdot\Omega_{\bl,n} &= 0, \cr}
\eqno(3.13)
$$
where the eigenvalues are in the ring $\BL$.
(That is, we may regard $V(\bl;\ba)$ as a free module of finite rank
over the ring $\BL$.)
Let $H$ be the subalgebra of $\Up$ generated by
$\psi_k$, $\phi_{-k}$, $k\in\BZ_{\ge0}$, and
let $H_0$ be the quotient of $H$ by its center.
We can extend the function $\chi_{\bl;\ba}$ to a homomorphism of
commutative algebras, $\chi_{\bl;\ba}:H_0\longrightarrow \BL$, which
preserves the grading.
(Strictly speaking we should write
$\chi_{\bl;\ba}(p(\psi_k))$, etc.,
$p$ being the canonical projection to $H_0$, but the above notation
is clear.)
This leads naturally to:
\defn{3.4}.
The character $\chi_{\bl;\ba}$ of the loop module $V(\bl;\ba)$
is defined by the eigenvalues $\chi_{\bl;\ba}(\psi_m)$,
$\chi_{\bl;\ba}(\phi_{-m})$ extended to a homomorphism of commutative
algebras
$$
\chi_{\bl;\ba}:H_0\longrightarrow \BL.
\eqno(3.14)
$$
\enddef
For $V(l;a)$, the eigenvalue formulae \refeq{3.13} have the explicit
form
$$
\eqalign{
\psi_0\cdot(\st^n\otimes v^l_0)
&= q^l(\st^n\otimes v^l_0) , \cr
\phi_0\cdot(\st^n\otimes v^l_0)
&= q^{-l}(\st^n\otimes v^l_0), \cr
\psi_m\cdot(\st^n\otimes v^l_0)
&= (q^l-q^{-l})(aq^l)^m(\st^{m+n}\otimes v^l_0), \cr
\phi_{-m}\cdot(\st^n\otimes v^l_0)
&= (q^{-l}-q^l)(aq^l)^{-m}(\st^{-m+n}\otimes v^l_0). \cr}
\eqno(3.13a)
$$
Further formulae which we shall need are
$$
\eqalign{
\psi(u)\cdot(\st^n\otimes v^l_0)
&= \Big({q^l-au\st\over1-aq^lu\st}\Big)
(\st^n\otimes v^l_0). \cr
\phi(u)\cdot(\st^n\otimes v^l_0)
&= \Big({q^{-l}-(au\st)^{-1}\over1-(aq^lu\st)^{-1}}\Big)
(\st^n\otimes v^l_0). \cr} \eqno(3.13b)
$$
These are, of course, formal expansions in the spirit of
\refeq{2.12}.
It follows that
$$
h_m\cdot(\st^n\otimes v^l_0)
=m^{-1}[lm]a^m(\st^{m+n}\otimes v^l_0)
\eqno(3.13c)
$$

{}From \refeq{2.9} we know that
$\Delta(\psi_k)\cdot\Omega_{\bl,n}=\sum_{j=0}^k(\psi_j\cdot
v^{l_1}_0) \otimes(\psi_{k-j}\cdot\Omega_{\bl^\prime,n})$, where
$\bl^\prime=l_2,\cdots,l_k$ and $\ba^\prime=a_2,\cdots,a_k$.
Moreover, this may be extended to the $k$-fold iteration of the
co-product.
This convolution property enables general formulae for the eigenvalues
to be constructed quite readily, either as explicit formulae,
or by using the multiplicative property which is implied, namely
$$
\eqalign{
\psi(u)\cdot\Omega_{\bl,n}&=\Bigg(\prod_{j=1}^k
{q^{l_j}-a_ju\st\over1-a_jq^{l_j}u\st}
\Bigg)\Omega_{\bl,n}, \cr
\phi(u)\cdot\Omega_{\bl,n}&=\Bigg(\prod_{j=1}^k
{q^{-l_j}-(a_ju\st)^{-1}\over1-(a_jq^{l_j}u\st)^{-1}}
\Bigg)\Omega_{\bl,n}. \cr}
\eqno(3.14b)
$$
\lem{3.2}.
Let $V(\bl;\ba)$ be a quantum loop module.
Then the character function has the additive property
$\chi_{\bl;\ba}(h_m)=\sum_{i=1}^k\chi_{l_i;a_i}(h_m)$,
$(m\in\BZ_{\not=0})$.
\endlem
\proof
This follows immediately from equations \refeq{2.12}
and \refeq{3.14b}. \qed

We proceed to an important theorem concerning the character
function.
\thm{3.1}.
The image of $\chi_{\bl;\ba}$ is a Laurent subring of $\BL$ \ie,
$\chi_{\bl;\ba}(H_0)=\BC[\st^r,\st^{-r}]$, for some integer $r>0$.
\endthm
\proof
{}From \refeq{3.13c} we find
$$
\chi_{\bl;\ba}(h_m)=m^{-1}(q-q^{-1})^{-1}\st^m
\Big(\sum_{j=1}^k(a_jq^{l_j})^m-
\sum_{j=1}^k(a_jq^{-l_j})^m\Big).
\eqno()
$$
Let $y_i$, $i=1,\ldots,p$ be the distinct elements of the set
$\{a_jq^{l_j}, \, -a_jq^{-l_j} \, | \, 1\leq j\leq k\}$,
and let $\mu_i$ be the multiplicity of $y_i$, then
$$
\chi_{\bl;\ba}(h_{2m-1})=(2m-1)^{-1}(q-q^{-1})^{-1}\st^{2m-1}
\sum_{i=1}^{p}\mu_i y^{2m-1}_i.
\eqno()
$$
The determinant of the matrix $(y^{2m-1}_i)_{1\leq i,m\leq p}$
is proportional to a Vandermonde determinant,
so that the characters $\chi_{\bl;\ba}(h_{2m-1})$ cannot all
vanish simultaneously in the range $1\leq m\leq p$.
Similarly, $\chi_{\bl;\ba}(h_{-2m+1})$
do not all vanish simultaneously in the same range.
Therefore there exists integers $r,s\in\{1,\ldots,2p-1\}$ such
that $\chi_{\bl;\ba}(h_r)\neq 0$ and $\chi_{\bl;\ba}(h_{-s})\neq 0$.
The rest of the proof proceeds exactly as in Lemma 4.1, p. 328
of \refto{Ch86}.
\qed
In the sequel we shall assume that that $r=1$, \ie\ that the map
$\chi_{\bl;\ba}$ is surjective, unless otherwise stated.

\subsection{A determinant formula.}
Our strategy is as follows.
First we shall prove, in \reflem{3.3c},
that the module $V(\bl;\ba)$, with the dimensions
of the factors in the order $l_1\le\ldots\le l_k$, is generated by
$\Omega_{\bl,n}$ provided that
$$
a_j/a_i\not=q^{(l_i+l_j-2p+2)},\qquad 0<p\le l_i,\qquad i<j.
\eqno(3.51)
$$
The dual module $V^*(\bl;\ba)$ is isomorphic to
$V(\bl^\prime;\ba^\prime)$, with $(\bl^\prime;\ba^\prime)$ obtained
from $(\bl;\ba)$ by reversal of order: the proof used in
\reflems{3.3b}{3.3c} may be modified to show that
$V(\bl^\prime;\ba^\prime)$ is is generated by $\Omega_{\bl^\prime,n}$,
but now under the conditions
$$
a_j/a_i\not=q^{-(l_i+l_j-2p+2)},\qquad 0<p\le l_j,\qquad i<j.
\eqno(3.52)
$$
Duality and isomorphism of the two modules so generated leads to their
irreducibility, and then further to the fact that all
$V(\bl^{\prime\prime};\ba^{\prime\prime})$, where
$(\bl^{\prime\prime};\ba^{\prime\prime})$ is obtained from
$(\bl;\ba)$ by arbitrary permutation, are irreducible and generated
by $\Omega_{\bl^{\prime\prime},n}$ (\refthm{3.4}).
Finally we show that, when the conditions of \refthm{3.4} are not
satisfied, the modules are reducible (\refthm{3.5}).

We recall some definitions and results from \refto{CP91} which
generalize the conditions \refeq{3.51,3.52} to arbitrary $(\bl;\ba)$.
\defn{3.5}.
Given an $(m+1)$-dimensional irreducible $\U$ module $V_m$, and a
parameter $a\in\CX$, the associated $q$-string $S_m(a)$ is the set
$S_m(a)=\{a^{-1}q^{m-1},\cdots,$ $a^{-1}q^{1-m}\}$.
\enddef
\defn{3.7}.
The $q$-strings $S_1$ and $S_2$ are said to be in general position if
either $S_1\cup S_2$ is not a $q$-string, or
$S_1\subset S_2$ or $S_2\subset S_1$.
\enddef
\lem{3.3}.
The $q$-strings $S_m(a)$ and $S_n(b)$ are in general position if and
only if $\{b/a\not=q^{\pm(m+n-2p+2)}\mid0<p\le\min(m,n)\}$.
\endlem
We shall show that the necessary and sufficient condition
for a loop module to be generated by $\Omega_{\bl,n}$
and irreducible is that all the $q$-strings are in general position.

As a preliminary to \reflem{3.3c}, we prove a result
concerning the coefficient matrix $A$ which appears in
equation \refeq{3.22} below.
This matrix has the definition
$$
A_{r,j}=\sum_{s=1}^r b_j^s\,d_{r-s,j+1},\qquad
1\le r\le k,\qquad
1\le j\le k,
\eqno(3.53)
$$
where $b_1=q^{l_1-2i}a_1$, ($0\le i<l_1$), $b_j=q^{l_j}a_j$,
($j>1$) and the $d_{s,j}$ are numerical factors in the character
formula $\psi_s\cdot(v^{l_j}_0\otimes\cdots\otimes v^{l_k}_0)=
d_{s,j}(\st^s\otimes v^{l_j}_0\otimes\cdots\otimes v^{l_k}_0)$,
together with $d_{s,k+1}=\delta_{s,0}$.
The columns of $A$ are discrete convolutions, as are the $d_{s,j}$.
So it is natural to introduce generating functions for the columns:
${\cal A}_j(u)=\sum_{r=1}^\infty A_{r,j} u^{r-1}$,
since they factorise:
$$
\eqalign{
{\cal A}_j(u)&=\Big(\sum_{r\ge0}^\infty b_j^{r+1}u^r\Big)
\Big(\sum_{r\ge0}^\infty d_{r,j+1}u^r\Big) \cr
&={b_j\over1-ub_j} \prod_{i=j+1}^k
{q^{l_i}-uq^{-l_i}b_i\over1-ub_i}.
 \cr}\eqno(3.54a)
$$
In this calculation we have used \refeq{3.13b}.
We use this to evaluate the determinant of $A$ in an elementary way.
\lem{3.3b}.
$$
\det A=\prod_r b_r
\prod_{r<s}(q^{l_s}b_r-q^{-l_s}b_s)
\eqno(3.55)
$$
\endlem
\proof
The determinant is a multinomial in the variables $b_j$: for $q=1$, it
is a Vandermonde determinant with the stated factorisation.
{}From \refeq{3.54a} we can see that the highest and lowest powers of
$q$ which can occur are correctly given by \refeq{3.55}.
Therefore \refeq{3.55} is proved if we can show that the determinant
has the stated linear factors $(q^{l_s}b_r-q^{-l_s}b_s)$, $(r<s)$,
since no further factors involving $q$ are then possible.
To obtain these factors, we show that if $b_r=q^{-2l_s}b_s$, there
exist constants $c_r,\ldots,c_s$, not all zero, such that
$\sum_{i=r}^s c_i{\cal A}_i(u)\equiv0$, which implies that the
columns of the matrix are linearly dependent for any $k$.
After substituting and simplifying, these equations read
$$
\sum_{i=r}^{s-1} {c_ib_i\over 1-ub_i}
\prod_{j=i+1}^s {q^{l_j}-uq^{-l_j}b_j\over1-ub_j}+
{c_sb_s\over1-ub_s}\equiv0.
\eqno(3.56)
$$
If $b_r=q^{-2l_s}b_s$, there is a cancellation in the first term,
which becomes
$$
{c_rb_r\over 1-ub_s}
\prod_{j=r+1}^{s-1} {q^{l_j}-uq^{-l_j}b_j\over1-ub_j}.
\eqno(3.57)
$$
There is a common factor $(1-ub_s)$ in the denominators in
\refeq{3.56}, which we discard; the other factors are $(1-ub_i)$,
$(r<i<s)$. Rationalising, we obtain $\sum_{i=r}^s c_i{\cal
P}_i(u)\equiv0$. But the polynomials ${\cal P}_i(u)$ are all of
degree $s-r-1$, and $s-r+1$ such polynomials are necessarily
linearly dependent.
\qed
\section{Reducibility and Irreducibility}

\subsection{Cyclic property.}
The following result is of central importance.
\lem{3.3c}.
Suppose that $(\bl,\ba)$ is such that the $q$-strings $S_{l_i}(a_i)$
are all in general  position relative to each other, and that the
dimensions of the terms in the tensor product are in one of the orders
$l_1\le\ldots\le l_k$ or $l_1\ge\ldots\ge l_k$.
Then the loop module $V(\bl;\ba)$ is generated by any one of
the following vectors:
$$
\Omega_{\bl,n}=
\st^n\otimes v^{l_1}_0\otimes\cdots\otimes v^{l_k}_0,
\eqno(3.15)
$$
where the $v^{l_i}_0$ are highest weight vectors of the irreducible
$\U$ modules $V_{l_i}$.
\endlem
\proof
We shall give the proof for the first ordering only.
It is enough to show that all the vectors
$$
\st^m\otimes y_1 v^{l_1}_0\otimes\cdots\otimes y_k v^{l_k}_0, \qquad
(y_i\in N_{-}),
\eqno(3.16)
$$
are in the submodule generated by $\Omega_{\bl,n}$.
By \refthm{3.1} and our assumption that $r=1$, we know that for any
$m,n\in\BZ$ there exists a $Q_{m-n}\in H_0$ such that
$\chi_{\bl;\ba}(Q_{m-n})=\st^{m-n}$, giving
$$
Q_{m-n}\cdot\Omega_{\bl,n}=\Omega_{\bl,m},\qquad(m\in\BZ).
\eqno(3.17)
$$
Therefore we have only to show that any one of the vectors
\refeq{3.16} is in the submodule generated by $\Omega_{\bl,n}$,
$(n\in\BZ)$.
We prove this by induction on $k$.
The case $k=1$ is obvious from the definitions.
Let $\bl^\prime=(l_2,\ldots,l_k)$,
$\ba^\prime=(a_2,\ldots,a_k)$ and define
$$
\Omega_{\bl^\prime,n}=
\st^n\otimes v^{l_2}_0\otimes\cdots\otimes v^{l_k}_0,
\eqno(3.18)
$$

First we prove that $V(\bl,\ba)$ is generated by
$v^{l_1}_{l_1}\otimes\Omega_{\bl^\prime,n}$ where
$v^{l_1}_{l_1}$ is the lowest weight vector in $V_{l_1}$.
By the induction hypothesis, $V(\bl^\prime,\ba^\prime)$ is
generated by $\Omega_{\bl^\prime,n}$.
Recall that $\Uq=N_{-}HN_{+}$.
For any $w\in V(\bl^\prime,\ba^\prime)$
there exists $x\in N_{-}$ such that $w=x\cdot\Omega_{\bl^\prime,n}$.
Using \refeq{2.8}:
$$
\Delta(x)\cdot(v^{l_1}_{l_1}\otimes\Omega_{\bl^\prime,n})=
v^{l_1}_{l_1}\otimes w.
\eqno(3.19)
$$
This means that $v^{l_1}_{l_1}\otimes V(\bl^\prime,\ba^\prime)$
is generated by $v^{l_1}_{l_1}\otimes\Omega_{\bl^\prime,n}$.
Now
$$
x^{+}_0\cdot(v^{l_1}_{l_1}\otimes w)=
(x^{+}_0\cdot v^{l_1}_{l_1})\otimes w
+(K\cdot v^{l_1}_{l_1})\otimes(x^{+}_0 w).
\eqno(3.20)
$$
So $v^{l_1}_{l_1-1}\otimes V(\bl^\prime,\ba^\prime)$ is
also generated from $v^{l_1}_{l_1}\otimes\Omega_{\bl^\prime,n}$.
A finite induction completes the first step of the proof.

The second step consists in showing that for each $(0\le i<l_1)$,
the vector $v^{l_1}_{i+1}\otimes\Omega_{\bl^\prime,n}$ is
generated by the vectors $v^{l_1}_i\otimes\Omega_{\bl^\prime,n+r}$,
for $(1\le r\le k)$.
Repeating this process, we will show that
$v^{l_1}_{l_1}\otimes\Omega_{\bl^\prime,n}$ is in the submodule
generated by  $v^{l_1}_0\otimes\Omega_{\bl^\prime,n}$.
{}From the first step and \refeq{3.17}, this will complete the proof.
Now
$$
x^{-}_r\cdot(\st^m\otimes v^{l_1}_i\otimes\Omega_{\bl^\prime})
=\st^m\otimes v^{l_1}_i\otimes(x^{-}_r\cdot\Omega_{\bl^\prime})+
\sum_{s=1}^r\st^m\otimes
(x^{-}_s\cdot v^{l_1}_i)\otimes(\psi_{r-s}\cdot\Omega_{\bl^\prime}).
\eqno(3.21)
$$
Iterating this process, we get
$$
x^{-}_r\cdot(\st^m\otimes v^{l_1}_i\otimes\Omega_{\bl^\prime})=
\sum_{j=1}^k A_{rj} \st^{m+s}\otimes
v^{l_1}_i\otimes\cdots\otimes(f\cdot v^{l_j}_0)
\otimes\cdots\otimes v^{l_k}_0,
\eqno(3.22)
$$
where the $A_{rj}$ were defined in \refeq{3.53}.
{}From \reflem{3.3b} we conclude that the equations do indeed have a
solution under the stated conditions.

Finally, we should note that there is a double induction, starting
with the case that $\Omega_{\bl^\prime,n}=\st^n\otimes v^{l_2}_0$.
In the  course of the process, every one of the conditions \refeq{3.51}
are required - precisely half of the general position conditions.

The remaining conditions \refeq{3.52} are required to repeat the proof
with the other ordering $l_1\ge\ldots\ge l_k$.
The main difference is that one notes, from the decomposition
\refeq{3.32}, that generation from the highest component
$\Omega_{\bl,n}$ is
equivalent to generation from the lowest:
$\bar{\Omega}_{\bl,n}=\st^n\otimes
v^{l_1}_{l_1}\otimes\cdots\otimes v^{l_k}_{l_k}$.
The first step becomes a proof that $V(\bl,\ba)$ is generated by
$v^{l_1}_{0}\otimes\bar{\Omega}_{\bl^\prime,n}$, whilst in the
second step one replaces $x^{-}_r$ in equations \refeq{3.21} and
\refeq{3.22} by $x^{+}_{-r}$ to show that for each
$(0<i\le l_1)$,  the vector
$v^{l_1}_{i-1}\otimes\bar{\Omega}_{\bl^\prime,n}$ is a linear
combination of vectors
$v^{l_1}_i\otimes\bar{\Omega}_{\bl^\prime,n+r}$,
for $(1\le r\le k)$.
\qed

\subsection{Irreducibility.}
We know from $\U$ theory that
$$
V_m\otimes V_n=\oplus_{p=0}^{\min(m,n)} V_{m+n-2p}
\eqno(3.32)
$$
Since $\U$ is a subalgebra of $\Uq$, the loop modules have such a
$\U$ weight decomposition, with highest component $V_N$,
$N=\sum_{j=1}^{k}l_j$.
The simplest case, $V(1,1;a,b)$, is instructive.
We have the following:
\point(i)
If $a/b=q^2$ and $q^4\not=1$, then $\Omega=v^1_0\otimes v^1_0$
generates the whole module, but the vector
$\Omega_1=v^1_0\otimes v^1_1-qv^1_1\otimes v^1_0$ generates a
one-dimensional submodule.
\point(ii)
If $b/a=q^2$ and $q^4\not=1$, then $\Omega=v^1_0\otimes v^1_0$
generates a three-dimensional submodule, whilst the vector
$\Omega_1=v^1_0\otimes v^1_1-qv^1_1\otimes v^1_0$ generates the
whole module.
\point(iii)
$V^*(1,1;a,b)$ has a complementary structure in each case.
\point(iv)
The above illustrates why it is necessary to prove that the
modules $V(\bl;\ba)$ and $V(\bl^\prime;\ba^\prime)$ are both generated
from their highest component, even if $\bl^\prime=\bl$.

Obviously modules which satisfy the conditions of \reflem{3.3c} cannot
have proper submodules which contain the  highest component $V_N$ of
\refeq{3.32}.
This leads immediately to the following result.
\lem{3.5}.
Under the conditions of \reflem{3.3c}, the
modules $V(\bl,\ba)$ and $V^*(\bl,\ba)$ are irreducible.
\endlem
\proof
Suppose that $W$ is a proper submodule of $V(\bl,\ba)$, then its
annihilator $W^0$ is a proper submodule of $V^*(\bl,\ba)$ which does
contain the highest component.
This is not possible.
\qed
\thm{3.4}.
Suppose that $V(\bl;\ba)$ satisfies the conditions of \reflem{3.3c},
and that $\bl^\prime=\sigma(\bl)$, $\ba^\prime=\sigma(\ba)$, where
$\sigma$ is a permutation of $k$ objects.
Then $V(\bl^\prime;\ba^\prime)$ is irreducible and is generated by
any vector of the form \refeq{3.15}.
\endthm
\proof
We have that $V(\bl;\ba)$ is generated by $\Omega_{\bl,n}$, and is
irreducible.
Moreover, $\Omega_{\bl^\prime,n}$ generates a submodule of
$V(\bl^\prime;\ba^\prime)$.
Define a map
$V(\bl;\ba)\longrightarrow V(\bl^\prime;\ba^\prime)$
by $x\cdot\Omega_{\bl,n}\mapsto x\cdot\Omega_{\bl^\prime,n}$,
for all $x\in\Uq$.
This is obviously a homomorphism of $\Uq$ modules which is injective
because $V(\bl;\ba)$ is irreducible.
We must show that it is surjective.
This follows from the fact that it preserves the grading, and the image
of the finite-dimensional weight space at each grading level already has
dimension $\prod_{j=1}^{k}l_j$.
\qed

The fact that the module $V(\bl;\ba)$ is generated by its highest
component $\Omega_{\bl,n}$ does not guarantee irreducibility.
But the converse does hold.
Therefore, the method used to prove this last theorem gives a number
of corollaries:
\cor{3.1}.
Suppose that $V(\bl;\ba)$ is irreducible and that
$\bl^\prime=\sigma(\bl)$, $\ba^\prime=\sigma(\ba)$.
for some permutation $\sigma$.
Then $V(\bl^\prime;\ba^\prime)$ is irreducible.
\endcor
\cor{3.2}.
Suppose that $V(\bl;\ba)$ and $V(\bl^\prime;\ba^\prime)$ are
irreducible.
Then the following are equivalent:
\point(i)
the characters are proportional:
$\chi_{\bl,\ba}=c\chi_{\bl^\prime,\ba^\prime}$.
\point(ii)
$V(\bl;\ba)$ is isomorphic to $V(\bl^\prime;\ba^\prime)$.
\point(ii)
there is a permutation
$\sigma$ and a $s\in\CX$ such that
$\bl^\prime=\sigma(\bl)$ and $\ba^\prime=s\,\sigma(\ba)$.
\endcor

\subsection{Reducibility.}
Let us now show that the conditions of general position are
necessary  as well as sufficient for irreducibility.
The crucial step is to consider the module $V(m,n;a,b)$.
\lem{3.7}.
Let $\Omega^s_p=\st^s\otimes\Omega_p$, where $\Omega_p$ is the
$\U$ highest weight vector of the component $V_{m+n-2p}$ in
\refeq{3.32} and $\Omega^s_p$ is a  weight vector in $V(m,n;a,b)$.
Then we have $N_{+}\cdot\Omega^s_p=0$, and
$h_k\cdot\Omega^s_p=c\Omega^{s+k}_p$ $(\forall k\in\BZ)$,
if and only if $b/a=q^{m+n-2p+2}$, ($0<p\le\min(m,n)$).
\endlem
\proof
The proof involves a long computation.
We give details only in the case that $m\le n$.
Then, the $\U$ highest weight vector $\Omega_p$ is given by
$$
\Omega_p=\sum_{i=0}^p (-1)^i q^{i(m-i+1)}
[m-i]![n-p+i]! v^m_i\otimes v^n_{p-i}.
\eqno(3.33)
$$
By construction, we have $x^{+}_0\cdot\Omega_p^s=0$.
By explicit computation we will show that
$$
x^{+}_{\pm1}\cdot\Omega_p^s=0, \qquad
h_{\pm1}\cdot\Omega_p^s=c_\pm\Omega_p^{s\pm1},
\eqno(3.34)
$$
for some constants $c_\pm$, if and only if $b/a=q^{m+n-2p+2}$.
This done, it is an easy induction to extend the condition to all the
generators $x^{+}_k$, $(k\in\BZ)$ of $N_{+}$ using the commutators
$$
[h_{-1},x^+_k]=(q+q^{-1})x^+_{k-1}, \qquad
[h_1,x^+_k]=(q+q^{-1})x^+_{k+1}.
\eqno(3.35)
$$

By direct computation, one first shows that
$x^{+}_{-1}\cdot\Omega_p^s=0$ if and only if $b/a=q^{m+n-2p+2}$.
This uses the isomorphism \refeq{2.15} to find
$\Delta(x^{+}_{-1})=\Delta(K^{-1}\ce^{-1} f_0)=
\Delta(K^{-1}\ce^{-1})(f_0\otimes t_0^{-1}+1\otimes f_0)$.
Therefore,
$$
(\varphi_a\otimes\varphi_b)\Delta(x^{+}_{-1})=
(K^{-1}\otimes K^{-1})(\st^{-1}a^{-1}e\otimes K+\st^{-1}\otimes
b^{-1}e) \eqno(3.36)
$$
which allows the computation to be completed.
Now consider the vector $\tilde{\Omega}_p^s=h_{-1}\cdot\Omega_p^s$.
{}From $[h_{-1},x^{+}_0]=(q+q^{-1})x^{+}_{-1}$
we find that $e_1\cdot\tilde{\Omega}_p^s=0$, \ie, $\tilde{\Omega}_p^s$
is also a $\U$ highest weight vector of the same $\U$ weight as
$\Omega_p^s$.
Moreover, $d\cdot\tilde{\Omega}_p^s=(s-1)\tilde{\Omega}_p^s$, which
shows that $\tilde{\Omega}_p^s$ is proportional to $\Omega^{s-1}_p$.

We must also prove the analogous result for $k=1$.
First we need that $h_1\cdot\Omega_p^s=c\Omega_p^{s+1}$
if and only if $b/a=q^{m+n-2p+2}$.
This time we use $h_1=[x^{+}_0,x^{-}_1]K^{-1}\ce$ together with
\refeq{2.15} to find
$$
\eqalign{
(\varphi_a&\otimes\varphi_b)\Delta(h_1)=
(1-q^2)a\st fe\otimes1+
(1-q^2)b\st\otimes fe+ \cr
&(q^{-2}-q^2)a\st fK\otimes e+
a\st\left({K-K^{-1}\over q-q^{-1}}\right)\otimes1+
b\st\otimes\left({K-K^{-1}\over q-q^{-1}}\right). \cr}
\eqno(3.37)
$$
A long computation gives the desired result.
Since $[h_1,x^{+}_0]=(q+q^{-1})x^{+}_1$, this gives
$x^{+}_1\cdot\Omega_p^s=0$.
The second statement, $h_k\cdot\Omega^s_p=c\Omega^{s+k}_p$
$(\forall k\in\BZ)$, is now easily proved.
\qed

\lem{3.8}.
$V(m,n;a,b)$ is irreducible if and only if $S_m(a)$ and  $S_n(b)$ are
in general position.
\endlem
The ``if'' part is the subject of \refthm{3.4}.
{}From \reflem{3.7} $V(m,n;a,b)$ has a submodule generated by
$\Omega_p^s$ if $b/a=q^{m+n-2p+2}$, for $0<p\le\min(m,n)$.
Recall that $V^*(m,n;a,b)$ is isomorphic to $V(n,m;q^2b,q^2a)$, so it
has a submodule generated in the same way when  $a/b=q^{m+n-2p+2}$.
Together these possibilities exhaust the conditions of the lemma.
Moreover, $V(m,n;a,b)$ and $V^*(m,n;a,b)$ must be both irreducible
or both reducible.
\qed

\thm{3.5}.
$V(\bl;\ba)$ is irreducible if and only if all the $q$-strings
$S_{l_i}(a_i)$  are in general position relative to each other.
\endthm
\proof
Assume first that some pair $S_{l_i}(a_i)$, $S_{l_j}(a_j)$, $(i\not=j)$,
are not in general position, but that $V(\bl;\ba)$ is
irreducible. By \refthm{3.4} we can assume that $j=i+1$.
But then \reflem{3.8} shows that there is a proper submodule.
The sufficiency of the condition was already proved.
\qed

\subsection{Integrability.}
For completeness we also note:
\thm{3.6}.
Every irreducible loop module $V(\bl;\ba)$ is integrable.
\endthm
\proof
The argument in proposition  2.2 of \refto{CP86} can be
very easily generalized to the quantum case.
\qed

\subsection{Tensor Products.}
In the previous section we saw that $\Uq$ loop modules share many of
the properties with finite-dimensional $\Up$ modules, particularly in
the way the parameters determine their irreducibility.
Tensor products of irreducible finite-dimensional $\Up$ modules are
again irreducible in the generic case.
However, this is not true for loop modules.
Let $\bl=(l_1,\ldots,l_k)$, $\ba=(a_1,\ldots,a_k)$, and
$\bl^\prime=(l_1^\prime,\ldots,l^\prime_{k^\prime})$,
$\ba^\prime=(a_1^\prime,\ldots,a^\prime_{k^\prime})$. We put
$(\bl,\bl^\prime)=
(l_1,\ldots,l_k,l_1^\prime,\ldots,l^\prime_{k^\prime})$ and
$(\ba,\ba^\prime)=
(a_1,\ldots,a_k,a_1^\prime,\ldots,a^\prime_{k^\prime})$.
Define the linear map, for $s,s^\prime\in\CX$,
$$
p_{s,s^\prime}:V(\bl;\ba)\otimes V(\bl^\prime;\ba^\prime)
\longrightarrow V(\bl,\bl^\prime;s\ba,s^\prime\ba^\prime)
\eqno(4.1)
$$
by
$$
p_{s,s^\prime}:(\st^n\otimes w)\otimes(\st^{n^\prime}\otimes w^\prime)
\mapsto
s^n s^{\prime n^\prime}
\,\st^{n+n^\prime}
\otimes w\otimes w^\prime
\eqno(4.2)
$$
One checks readily that this is a $\Uq$-linear map, using the
coassociativity of the co-product.
In the generic case we may assume that
$V(\bl;\ba)$, $V(\bl^\prime;\ba^\prime)$ and
$V(\bl,\bl^\prime;s\ba,s^\prime\ba^\prime)$ are all irreducible.
Thus we see that $V(\bl;\ba)\otimes V(\bl^\prime;\ba^\prime)$ has
uncountably many quotients.
This is already noted, for the Lie algebra case, in
\refto{CP86}.

In view of \refcor{3.2}, we may restrict ourselves to
the case that $s^\prime=1$.
We also define the algebraic direct integral,
$$
\int^\oplus V(\bl,\bl^\prime;s\ba,\ba^\prime)\,ds,
\eqno(4.3)
$$
as in \refto{CP86}, to be the space of algebraic maps
$$
\omega:\CX\longrightarrow
\BL(\mathop{\otimes}\limits_{i=1}^k V_{l_i}
\mathop{\otimes}\limits_{i=1}^{k^\prime} V_{l^\prime_j}),
\eqno(4.4)
$$
equipped with the $\Uq$ action given by
$$
(x\cdot w)(s)=P_\BL\circ
(\hat{\pi}_{\bl;s\ba}\otimes\hat{\pi}_{\bl^\prime;\ba^\prime})\circ
\Delta(x)\,w(s).
\eqno(4.5)
$$
\thm{4.1}.
$\int^\oplus V(\bl,\bl^\prime;s\ba,\ba^\prime)\,ds$
is isomorphic to $V(\bl;\ba)\otimes V(\bl^\prime;\ba^\prime)$.
\endthm
\proof
Assign to each
$(\st^n\otimes w)\otimes(\st^{n^\prime}\otimes w^\prime)$ the map
given by
$\omega(s)=s^n\st^{n+n^\prime}\otimes w\otimes w^\prime$.
This map is $\Uq$-linear, and its kernel is obviously trivial.
Clearly the direct integral is generated by the map
$\omega_{\Omega,\Omega^\prime}(s)=s^n\Omega\otimes\Omega^\prime$,
where $\Omega$ and $\Omega^\prime$, with degree $n$, $n^\prime$, are
the generators of $V(\bl;\ba)$, $V(\bl^\prime;\ba^\prime)$
respectively.
Therefore the $\Uq$-linear map is also surjective.
\qed
\section{Crystal Base}

\subsection{Basic notions.}
We recall from \refto{Ka90} some basic notions concerning the upper
and lower crystal lattices and bases, here restricted to integrable
$\Uq$ modules.
They are related by an appropriate choice of coalgebra structure.
The co-product and antipode defined in \refeq{2.6} will be called
upper coproduct $\Delta_{+}$ and antipode $\ap_{+}$.
The lower coproduct $\Delta=\Delta_-$ and the corresponding
antipode $\ap_-$ are defined by
$$
\eqalign{
&\Delta_{-}(e_i)=e_i\otimes t_i^{-1}+1\otimes e_i, \cr
&\Delta_{-}(f_i)=f_i\otimes 1+ t_i\otimes f_i, \cr
&\Delta_{-}(t_i)=t_i\otimes t_i, \cr}\qquad
\eqalign{
&\ap_{-}(e_i)=-e_it_i, \cr
&\ap_{-}(f_i)=-t_i^{-1}f_i, \cr
&\ap_{-}(t_i)=t_i^{-1}. \cr}
\eqno(5.1)
$$
To define crystal lattices and bases one
considers $\Uq$ as an algebra over the rational function field
$\BQ(q)$, where $q$ is now a formal variable,
or a transcendental number over $\BQ$ \refto{Ka90}.

Let $M$ be an integrable $\Uq$ module
with finite-dimensional weight spaces $M_{\nu}$,
and let $A(q)$ be the subring of
$\BQ(q)$ consisting of rational functions regular at $q=0$.
We describe the construction of the lower crystal lattice/base.
For  $i=0$, $1$, any weight vector $u\in M$ belongs to a $\U$
subalgebra  generated by $e_i$, $f_i$, and $M$
can therefore be uniquely
decomposed as
$$
M = \bigoplus_{0\leq n\leq\langle h_i,\nu\rangle}
    f_i^{(n)}\cdot(\Ker e_i \cap M_{\nu}),
\eqno()
$$
where $e_i^{(k)}=e_i^k/[k]!$, $f_i^{(k)}=f_i^k/[k]!$.
One defines linear maps $\et_i$, $\ft_i$,
in $\End(M)$ as follows \refto{Ka90}.
$$
\ft_i\cdot(f_i^{(n)}\cdot u) = f_i^{(n+1)}\cdot u, \qquad
\et_i\cdot(f_i^{(n)}\cdot u) = f_i^{(n-1)}\cdot u,
\eqno(5.4)
$$
for $u\in\Ker e_i \cap M_{\nu}$ with
$0\leq n\leq\langle h_i,\nu\rangle$.
There is a similar definition in \refto{Ka90} of
operators $\et_i^{up}$ and $\ft_i^{up}$ suitable for
the upper crystal base.
The crystal lattice and crystal base at  $q=0$ are defined using
these modified Chevalley generators $\et_i$, $\ft_i$.
The pair $(L,B)$ is a lower crystal base of the integrable module $M$
if and only if the following properties hold.
\point(i)
$L$ is a free $A(q)$\/-module such that $M=L\otimes_{A(q)}\BQ(q)$.
\point(ii)
$B$ is a base of the $\BQ$-vector space $L/qL$.
\point(iii)
$\et_{i}L\subset L$, $\ft_{i}L\subset L$ and
$\et_{i}B\subset B\cup\{0\}$, $\ft_{i}B\subset B\cup\{0\}$.
\point(iv)
Corresponding to the weight space decomposition $M=\oplus M_\nu$ we
have $L=\oplus L_\nu$ and $B=\cup B_\nu$ with $L_\nu=L\cap M_\nu$
and $B_\nu=B\cap M_\nu$.
\point(v)
If $b$, $b^\prime\in B$ then $b^\prime=\ft_i\cdot b$ if and only if
$b=\et_i\cdot b^\prime$.

$L$ is called the crystal lattice, and $B$ is the crystal base.
The properties of a crystal base $(L,B)$ of an integrable $\Uq$
module $M$, when it exists, are captured
in the definition of the crystal graph ${\cal G}$.
The vertices of ${\cal G}$ are the elements of $B$.
To each pair $b,b'\in B$ such that $b'=\ft_i \cdot b$, there
corresponds an arrow from $b$ to $b'$, labeled by $i$.

An integrable highest weight module $V(\lambda)$ has the
standard crystal base at $q=0$ described as follows.
$$
\eqalignno{
&L(\lambda)=
\sum A(q)\ft_{i_1}\cdots \ft_{i_k}\cdot u_\lambda,
\qquad
B(\lambda)=L(\lambda)\mod qL(\lambda)\backslash\{0\},
&(5.5:a) \cr
&L^{up}(\lambda)_\nu=
 q^{(\lambda,\lambda)-(\nu,\nu)}L(\lambda)_\nu, \qquad\
B^{up}(\lambda)_\nu=
 q^{(\lambda,\lambda)-(\nu,\nu)}B(\lambda)_\nu.
&(5.5:b) \cr}
$$
It is easy to see that the crystal base of the $U_q(sl_2)$
module $V_l$ defined in \refeq{2.0a} is given by
$v_k^l$, $k=0,\ldots,l$.

\subsection{Construction for loop modules.}
We shall now present some results on the crystal bases
of loop modules. We shall see that not all the loop
modules $V(\bl;\ba)$ admit a crystal base.
Furthermore, when a crystal
base exists, it may not be unique, and the crystal graph is not
always connected
(recall that uniqueness of $B$ is always $\mod qL$). 
We shall consider only the lower crystal base,
the upper one is related to the lower one by the relations such as
those in \refeq{5.5:b} above \refto{Ka90,DJO92}.

We start with two basic observations.
\lem{5.0}.
\point(i)
Suppose the loop module $V(\bl;\ba)$ has a crystal base $(L,B)$.
Then every $b\in B$ is proportional {\rm(}$\mod qL${\rm)} 
to some vector of the
form $\st^p\otimes v^{l_1}_{j_1}\otimes\cdots\otimes v^{l_k}_{j_k}$,
where $p\in\BZ$ and $0\leq j_i\leq l_i$, $i=1,\ldots,k$.
\point(ii)
Assume $(L,B)$ and $(L',B')$ are
two crystal bases of a loop module $V(\bl;\ba)$, with $L=L'$.
Let ${\cal S}$ be a connected subgraph of the crystal graph
${\cal G}$, and let $S$ and $S'$ be the subsets of $B$ and $B'$
corresponding to ${\cal S}$. Then $S'=\alpha S$
{\rm(}$\mod qL${\rm)}, $\alpha\in\BQ$. 
\endlem
\proof
Let $U_1$ be the $U_q(sl_2)$ subalgebra of $\Uq$
generated by $e_1$ and $f_1$. The structure of the module
$V(\bl;\ba)$ restricted to $U_1\oplus\BC d$ is as follows:
$$
V(\bl;\ba) = \bigoplus_{p\in\BZ} M_p,
\eqno()
$$
where $M_p$ is the eigenspace of $d$
with eigenvalue $p$ and
$$
M_p \cong M=V_{l_1}\otimes\cdots\otimes V_{l_k},
\eqno()
$$
as a $U_q(sl_2)$ module.
We know by proposition 6 of \refto{Ka90}
that the crystal base of $M$
may be taken as 
$B^{(1)}=\{ v^{l_1}_{j_1}\otimes\cdots\otimes v^{l_k}_{j_k}\}$.
By the above-mentioned point (iv) of the definition of a crystal base,
$B$ must have a decomposition:
$$
B = \bigcup_{p\in\BZ} B_p,
\eqno(5.0)
$$
with $B_p$ the subset of eigenvectors of $d$ with eigenvalue $p$.
Now $B$ has to be a crystal base with respect to $U_1$, and
by the uniqueness theorem for crystal bases of finite-dimensional
representations of $U_q(sl_2)$ \refto{Ka90},
every $b\in B_p$ is proportional to some vector of
$\st^p\otimes B^{(1)}$ ($\mod qL$). 
This proves (i).

Suppose $(L,B)$ and $(L',B')$ are two crystal bases of
$V(\bl;\ba)$. Then by (i) every $b\in B$ is proportional to
some $b'\in B'$. Take $b_0\in S$, $b'_0\in S'$ such that
$b'_0=\alpha b_0$, $\alpha\in\BQ$.
By the assumption of connectedness,
every $b\in S$ can be written as $b=P\cdot b_0$, where $P$ is
a monomial in the variables $\et_i$, $\ft_i$, $i=0,1$.
Therefore, $\alpha b=P\cdot b'_0\in S'$, and (ii) is proved.
\qed

Next we note the following obvious, but important
\refto{KKMMNN92} 
\lem{5.01}.
Let $a\in\BQ$, and $L$ be the free $A(q)$\/-module generated by
$\st^p\otimes v^m_k$, $p\in\BZ$, $0\leq k\leq m$.
The loop module $V(m;a)$
has a crystal base $(L,B)$ given by
$B=\{\st^p\otimes a^p v^m_k$, $p\in\BZ$,
$0\leq k\leq m\}$. Its crystal graph is connected and the
crystal base $(L,B)$ is unique
{\rm(}$\mod qL${\rm)} 
up to an overall multiplicative constant.
\endlem
The crystal graph of $V(m;a)$ is given on figure 1.
The solid arrows
correspond to the action of $\ft_1$, the dashed ones to
$\ft_0$.

The remainder of this section is devoted to a study of $V(m,n;a,b)$.
We assume that $a,b\in\BQ$.
Because $q$ is transcendent over $\BQ$, this automatically
implies that $S_{m}(a)$ and $S_{n}(b)$ are in
general position, and thus that $V(m,n;a,b)$
is irreducible.
We begin with a simple remark. Suppose $f(q)\in \BQ(q)$,
and suppose the term in the Laurent expansion of $f(q)$ about $q=0$
with the lowest exponent of $q$ is $a q^N$, where
$a\in\CX$, $N\in\BZ$. We abbreviate this by
$f(q) \approx a q^N$. Then we have, for $k\in\BZ_{\geq 0}$
$$
[k] \approx q^{-k+1}, \qquad [k]^{-1} \approx q^{k-1}.
\eqno()
$$

\figure{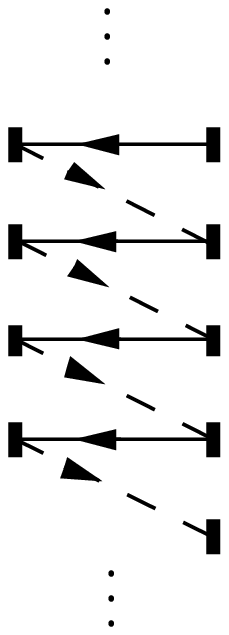}{Figure 1. The crystal graph of $V(1;a)$}

By proposition 6 of
\refto{Ka90}, we know that the crystal base of $V_m \otimes V_n$
may be taken as 
$B=\{v^m_j \otimes v^n_k\mid 0\leq j\leq m, 0\leq k\leq n\}$. 
Hence the crystal base of $V(m,n;a,b)$ considered as a
$U_q(sl_2)$ module is given by
$$
B_1 = \{\st^p \otimes v^m_j \otimes v^n_k \mid
0\leq j\leq m,\ 0\leq k\leq n,\ p\in\BZ \},
\eqno()
$$
and the action of $\ft_1$ on $B_1$ is:
$$
\eqalignno{
\ft_1\cdot(\st^p \otimes v^m_j \otimes v^n_k) &=
\st^p \otimes v^m_{j+1} \otimes v^n_k
\qquad\; \mod qL_1, \qquad m-j>k, &(5.6:a) \cr
\ft_1\cdot(\st^p \otimes v^m_j \otimes v^n_k) &=
\st^p \otimes v^m_j \otimes v^n_{k+1}
\qquad \mod qL_1, \qquad m-j\leq k. &(5.6:b) \cr}
$$
Denote by $L_1$ the vector space over $\BQ(q)$  spanned by $B_1$.
The generators $e_0$ and $f_0$ act on $V(m,n;a,b)$ as the
operators
$$
e_0 = \st \otimes(a f_1 \otimes t_1 + b \otimes f_1), \qquad
f_0 = \st^{-1}
\otimes  (a^{-1} e_1 \otimes 1 + b^{-1} t_1^{-1} \otimes e_1).
\eqno()
$$
\lem{5.1}.
Let $p\in\BZ$ and $k\in\{0,1,\ldots,{\rm min}(m,n)\}$, and define
$\omega^p_k = \st^p\otimes v^m_m \otimes v^n_{n-k}$.
Then
$$
V(m,n;a,b) \cap
\Ker e_0\ni \omega^p_k
\qquad \mod qL_1.
\eqno()
$$
\endlem
\proof
Consider the linear combination
$$
\Omega^p_k = \st^p\otimes \sum_{j=0}^{k} \alpha_j
v^m_{m-j} \otimes v^n_{n+j-k}.
\eqno()
$$
We obtain after some simple manipulations
$$
\eqalignno{
e_0\cdot\Omega^p_k &= \st^{p+1}\otimes &() \cr
&\sum_{j=1}^{k}
(\alpha_j a[m-j+1] q^{-n+2k-2j}+\alpha_{j-1} b[n+j-k])
v^m_{m-j+1} \otimes v^n_{n+j-k}. \cr}
$$
Thus $\Omega^p_k\in \Ker e_0$ if and only if
$$
\alpha_j a[m-j+1] q^{-n+2k-2j}+\alpha_{j-1} b[n+j-k] = 0,
\qquad\forall j.
\eqno()
$$
Therefore,
$$
\alpha_j = - {b\over a} {[n+j-k]\over [m-j+1]}
q^{n-2k+2j} \alpha_{j-1}.
\eqno()
$$
If we set $\alpha_0 = 1 =$ coefficient of
$\omega^p_k$ in $\Omega^p_k$, we get
$$
\alpha_j = (-1)^j \left( {b\over a} q^{n-2k}\right)^j
\prod_{l = 1}^j {[n+l-k]\over [m-l+1]} q^{2l},
\eqno()
$$
and we find $\alpha_j \approx {\rm const} \, q^{j(m-k+1)}$.
\qed
\lem{5.2}.
Let $p\in\BZ$, $0\leq j\leq m$ and $0\leq k\leq n$. Then
$$
\eqalignno{
\ft_0\cdot(\st^p v^m_j \otimes v^n_{n-k}) &=
\st^{p-1} a^{-1} v^m_{j-1} \otimes v^n_{n-k}
\qquad\; \mod qL_1, \qquad j>k, \qquad &(5.7:a) \cr
\ft_0\cdot(\st^p v^m_j \otimes v^n_{n-k}) &=
\st^{p-1} b^{-1} v^m_j \otimes v^n_{n-k-1}
\qquad \mod qL_1, \qquad j\leq k. &(5.7:b) \cr }
$$
\endlem
\proof
Let us start by showing \refeq{5.7:a}.
We have
$$
\eqalignno{
f_0&\cdot(\st^p\otimes v^m_j \otimes v^n_{n-k}) &(5.9) \cr
&= \st^{p-1}\otimes
(a^{-1}[m-j+1] v^m_{j-1} \otimes v^n_{n-k} +
b^{-1} q^{-m+2j} [k+1] v^m_j \otimes v^n_{n-k-1}). \cr}
$$
The proof goes by induction on $j$.
The case $j=m$ follows from the previous lemma.
Assume \refeq{5.7} has been proved for $j+1,j+2,\ldots,m$. Then
$$
a^{j-m}\st^p\otimes v^m_j \otimes v^n_{n-k} =
\ft_0^{m-j}\cdot\omega^{p+m-j}_k = f_0^{(m-j)}\cdot\omega^{p+m-j}_k
\qquad \mod qL_1.
\eqno()
$$
But \refeq{5.4} tells us that
$$
\ft_0\cdot(\st^p\otimes v^m_j \otimes v^n_{n-k}) = [m-j+1]^{-1}
f_0\cdot(\st^p\otimes v^m_j \otimes v^n_{n-k}),
\eqno()
$$
therefore the properly normalized second term of \refeq{5.9}
belongs to $qL_1$, since
$$
[m-j+1]^{-1} q^{-m+2j} [k+1] \approx q^{j-k}.
\eqno()
$$
Next we prove the case $j=k$ of \refeq{5.7:b}. Put
$$
\Omega_j = \st^p\otimes \sum_{k=0}^j (-1)^k
\left({b\over a}\right)^k q^k
v^m_{j-k} \otimes v^n_{n-j+k}.
\eqno()
$$
Note that
$\Omega_j = \st^p\otimes v^m_j \otimes v^n_{n-j} \mod qL_1$.
It is easy to see that
$$
\eqalign{
f_0&\cdot\Omega_j = \st^{p-1}\otimes b^{-1}q^{-m+2j}[j+1]v^m_j\otimes
v^n_{n-j-1} \cr
& - \st^{p-1}\otimes \sum_{k=1}^j (-1)^k
\left({b\over a}\right)^k q^k b^{-1}\beta_{jk}
v^m_{j-k} \otimes v^n_{n-j+k-1},\cr}
\eqno(5.10)
$$
where
$$
\beta_{jk} =
q^{-1}[m-j+k]-q^{-m+2j-2k}[j-k+1]
\approx q^{-m+j-k}-q^{-m+j-k} = 0.
\eqno()
$$
For the coefficient of the first term of \refeq{5.10} we find
$q^{-m+2j}[j+1] \approx q^{-m+j}$.
As above, \refeq{5.10} has to be normalized by multiplying it
with the factor $[m-j+1]^{-1} \approx q^{m-j}$. This proves that
$\ft_0\cdot\Omega_j = \st^{p-1}b^{-1}v^m_j\otimes v^n_{n-j-1}$
$(\mod qL_1)$.

It remains to prove \refeq{5.7:b} when $j<k$. Assume
it has been proved already for $k-1,k-2,\ldots,j$. The case $k=j$
has just been dealt with.
By the induction hypothesis,
$$
b^{-k+j-m+j}\st^p\otimes v^m_j \otimes v^n_{n-k} =
f_0^{(k-j+m-j)}\cdot\omega^{p+m+k-2j}_k \qquad \mod qL_1.
\eqno()
$$
Hence the normalizing factor is $[m+k-2j+1]^{-1}$.
Consider the two terms on the r.h.s. of \refeq{5.9}. The coefficient
of the first one becomes
$[m+k-2j+1]^{-1}[m-j+1] \approx q^{k-j}$, and the second one
$[m+k-2j+1]^{-1}q^{-m+2j}[k+1] \approx q^0$.
\qed

We can summarize the previous two lemmas as follows.
Put $v^p_{jk}=\st^p\otimes v^m_j\otimes v^n_k$.
Then the action of the modified Chevalley generators on this base
can be read from \refeq{5.6,5.7}:
$$
\eqalignno{
\ft_0\cdot v^p_{jk} &= a^{-1} v^{p-1}_{j-1,k}
\qquad j>n-k, &(5.11:a) \cr
\ft_0\cdot v^p_{jk} &= b^{-1} v^{p-1}_{j,k-1}
\qquad j\leq n-k. &(5.11:b) \cr
\ft_1\cdot v^p_{jk} &= v^p_{j+1,k}
\qquad m-j>k, &(5.12:a) \cr
\ft_1\cdot v^p_{jk} &= v^p_{j,k+1}
\qquad m-j\leq k. &(5.12:b) \cr}
$$
Because there are structure constants different from $1$, we call
it the pseudo-crystal base. Its importance stems from the fact that
each vector of any crystal base must be proportional to some
$v^p_{jk}$ by \reflem{5.0}(i).
We also define in the obvious way the notion
of pseudo-crystal graph.
To illustrate this, we give
in figure 2 the pseudo-crystal graph of $V(3,2;a,b)$. The
graph itself can be thought of as being 3-dimensional, the
two horizontal dimensions corresponding to the indices
$j,k$, and the vertical dimension to $p$. The figure is the
projection of the graph on the $p=0$ plane. The solid arrows
correspond to the action of $\ft_1$, the dashed ones to
$\ft_0$, which has of course also a vertical component not shown
here.

\defn{5.1}.
Let $j,k\in\BZ$ be such that $0\leq j\leq m$, $0\leq k\leq n$.
We say that $V(m,n;a,b)$ has an escalator at $(j,k)$,
if for every $p\in\BZ$, $\ft_0\ft_1\cdot v^p_{jk}=c\, v^{p-1}_{jk}$,
where $c$ is either $a^{-1}$ or $b^{-1}$.
If there exists $(j,k)$ such that $V(m,n;a,b)$ has an escalator at
$(j,k)$, but we do not want to specify the value(s) of $(j,k)$,
then we just say that $V(m,n;a,b)$ has an escalator.
\enddef
\figure{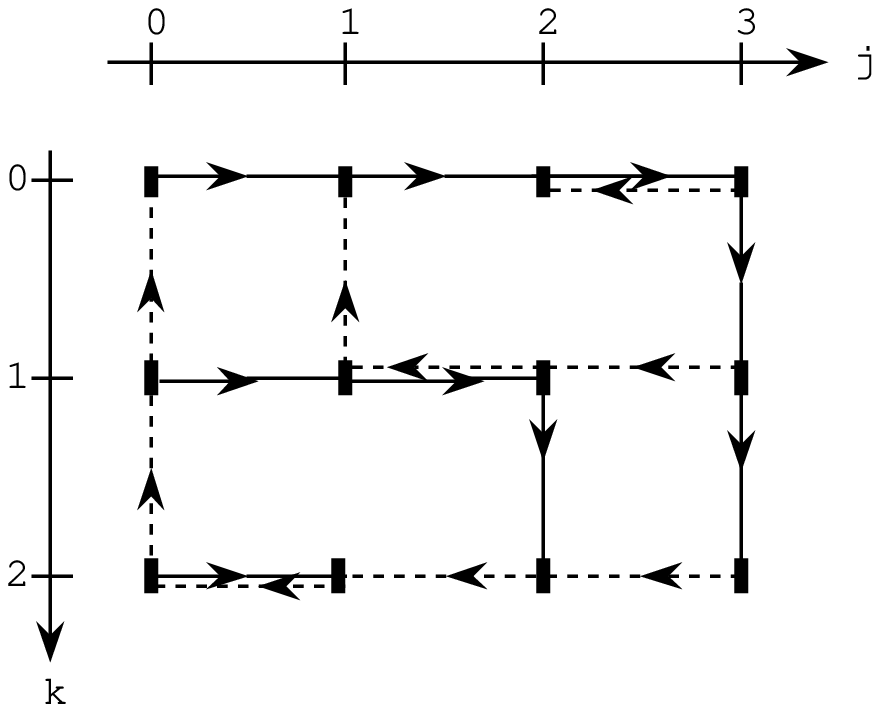}{Figure 2. $V(3,2;a,b)$} 
\lem{5.3}.
$V(m,n;a,b)$ has an escalator if and only if $m\not=n$.
More precisely: if $m>n$, for every $k\in\{0,\ldots,n\}$,
there exists $j$ such that
$V(m,n;a,b)$ has an escalator at $(j,k)$, and if $m<n$,
for every $j\in\{0,\ldots,m\}$,
there exists $k$ such that
$V(m,n;a,b)$ has an escalator at $(j,k)$.
\endlem
\proof
There are two possibilities for an escalator: either
$$
\eqalignno{
\ft_1\cdot v^p_{jk} &= v^p_{j+1,k}\qquad m-j>k
&(5.15:a)\cr
\ft_0\cdot v^p_{j+1,k} &= a^{-1} v^{p-1}_{jk}\qquad j+1>n-k
&(5.15:b)\cr}
$$
or
$$
\eqalignno{
\ft_1\cdot v^p_{jk} &= v^p_{j,k+1}\qquad m-j\leq k
&(5.16:a)\cr
\ft_0\cdot v^p_{j,k+1} &= b^{-1} v^{p-1}_{jk}\qquad j\leq n-k-1
&(5.16:b)\cr}
$$
In order for the first one, \refeq{5.15},
to be realized, we must have $n-k-1<j<m-k$, so that $m>n$.
For the second one, \refeq{5.16},
we must have $m-j\leq k\leq n-j-1$, hence $m<n$.
Conversely, if $m\not=n$, it is clear that we can find
$j,k$ satisfying the conditions in \refeq{5.15} or \refeq{5.16}.
\qed
\figure{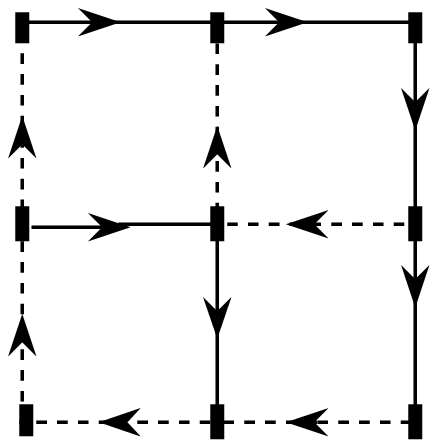}{Figure 3a. $V(2,2;a,b)$} 
\figure{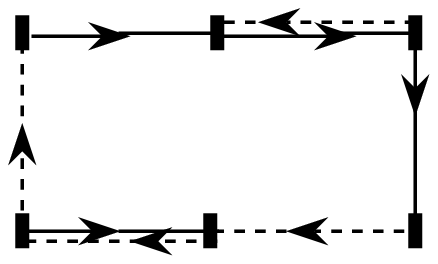}{Figure 3b. $V(2,1;a,b)$}
\figure{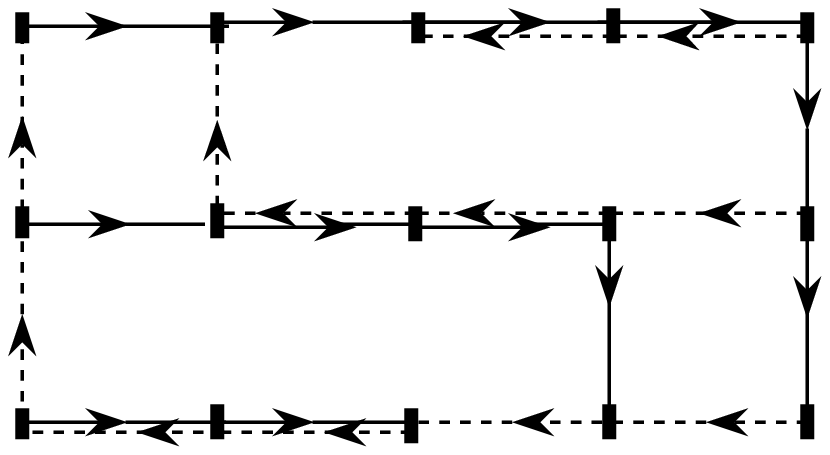}{Figure 3c. $V(4,2;a,b)$}

The escalators are easy to recognize on the projection
of the pseudo-crystal graph. They correspond to the pairs of vertices
in the projection which are connected by two arrows of opposite
direction, one coming from $\ft_1$ and the other from $\ft_0$.
In figure 3,
we give the projections of the pseudo-crystal graphs
of $V(2,2;a,b)$, $V(2,1;a,b)$ and $V(4,2;a,b)$.

For $0\leq j\leq m$, $0\leq k\leq n$, we define the quantity
$d_{jk}={\rm min}(k,m-j)$.
Then for each $p\in\BZ$, $N\in\{0,\ldots,{\rm min}(m,n)\}$,
the vectors $v^p_{jk}$
such that $d_{jk}=N$ form the crystal base of an irreducible
submodule of the algebra generated by $e_1$ and $f_1$.

\thm{1}. 
Let $L$ be the free $A(q)$\/-module generated by $v^p_{jk}$,
with $p\in\BZ$, $0\leq j\leq m$, $0\leq k\leq n$.
\point(i)
If $m=n$ and $\sqrt{b/a}\in\BQ$, then the vectors
$$
b^{p,\epsilon}_{jk} = \epsilon^{p+d_{jk}}
(ab)^{p/2} (b/a)^{d_{jk}/2} v^p_{jk},
\eqno()
$$
where $\epsilon=\pm 1$,
define two crystal bases $(L,B^{(+1)})$
and $(L,B^{(-1)})$ of $V(m,n;a,b)$, which are not proportional.
\point(ii)
If $m\not=n$, then there exists no crystal base of $V(m,n;a,b)$
unless $a=b$, in which case the vectors
$$
b^p_{jk} = a^p v^p_{jk},
\eqno()
$$
define a crystal base $(L,B)$, and every other crystal base
is proportional to $(L,B)$.
\endthm 
\proof
The remark about $d_{jk}$ preceding the theorem 
proves that
$B^{(\pm 1)}$ are stable under $\et_1$, $\ft_1$.
Assume $j>m-k$. Then
$$
\ft_0\cdot b^{p,\epsilon}_{jk} = \epsilon^{p+d_{jk}}
a^{-1} (ab)^{p/2} (b/a)^{d_{jk}/2}
v^{p-1}_{j-1,k},
\eqno(5.51)
$$
and $d_{j-1,k}=d_{jk}+1$. Indeed, $d_{j-1,k}=\min(k,m-j+1)$,
but $k<m-j+1$ together with $j>m-k$ implies $m-j<k<m-j+1$, which
is impossible for $k\in\BZ$. Thus $k\geq m-j+1>m-j$, and
$d_{j-1,k}=m-j+1=\min(k,m-j)+1=d_{jk}+1$. From this it follows
that the r.h.s. of \refeq{5.51} is equal to
$b^{p-1,\epsilon}_{j-1,k}$.
The case $j\leq m-k$ is similar. This proves (i).

Now we show (ii).
By \reflem{5.3}, $V(m,n;a,b)$ has $\min(m,n)$ escalators.
Consider the subgraph ${\cal S}_{jk}$
of the pseudo-crystal graph corresponding
to the escalator at $(j,k)$. This subgraph coincides with the crystal
graph of $V(1;a')$. Thus if $V(m,n;a,b)$ has a crystal base $(L,B)$,
with the decomposition \refeq{5.0} into eigenvectors of $d$,
and if $b\in B_p$ is a vertex of ${\cal S}_{jk}$, then
\reflem{5.0} and \reflem{5.01} imply
$$
b=\alpha_{jk}\, x^p v^p_{jk},
\eqno(5.50)
$$ where $x=a$ or $b$ according to whether
$m>n$ or $m<n$, and $\alpha_{jk}\in\BQ$ is independent of $p$.
Next, observe that \refeq{5.50} holds for every $b\in B_p$ such that
$b$ is proportional to $v^p_{j'k'}$ with $d_{j'k'}=d_{jk}$,
as $B$ must be crystalline for $\et_1$, $\ft_1$.
Hence there are constants $\alpha_i\in\BQ$, $i=1,\ldots,\min(m,n)$,
such that for every $b\in B_p$, $b=\alpha_i x^p v^p_{jk}$ with
$i=d_{jk}$.

Suppose now $m>n$,
and let us concentrate our attention on the last two rows
$k=n-1$, $n$ of the projection of the pseudo-crystal graph.
The reader may well look at the examples on the figures above,
which illustrate our arguments. The left side of these
two rows contains the two solid lines corresponding to the
vertices with $d_{jk}=n-1$ and $n$. They are connected
on the last row by a horizontal
dashed arrow corresponding to the action of $\ft_0$.
The effect of $\ft_0$ in this case is given by \refeq{5.11:a},
therefore $\alpha_n=\alpha_{n-1}$.
But they are also connected on the first column $j=0$ by
a vertical arrow, in which case the action of $\ft_0$ is
given by \refeq{5.11:b}, from which we deduce $a=b$.
An easy induction now shows that all the $\alpha_i$ are equal.
The case $m<n$ is similar.
\qed

\subsection{Connectedness of the crystal graph.}
In this paragraph we assume that
the loop module $V(m,n;a,b)$
is such that it has a crystal base $B=\{b^p_{jk}\}$, given in
\refthm{1}. 
Unlike the case of highest weight modules,
the crystal graph of $V(m,n;a,b)$ is not always
connected, as can be seen
for instance in the case of $V(1,1;a,b)$, where it has two
components. We shall now investigate this phenomenon in
greater detail. To begin, we need the following remark. Let
$M$ be an integrable $\Uq$ module, $v\in M$, then
$$
\et_i \cdot v = C_i(v,q) e_i \cdot v, \qquad
\ft_i \cdot v = D_i(v,q) f_i \cdot v, \qquad i=0,1,
\eqno(5.13)
$$
where $C_i(v,q),D_i(v,q)\in\BQ(q)$.
This follows from comparing \refeq{5.4} with \refeq{2.0a}.
Thus if $M$ has a crystal base $(L,B)$, $u\in B$, then
$\et_i \ft_j \cdot u=\epsilon\ft_j \et_i \cdot u$, where
$i\not=j$ and $\epsilon\in\{0,1\}$, in view of \refeq{2.1}.
Now if $\epsilon=0$, then \refeq{5.13} implies
$f_j\cdot u\in\Ker e_i$, and
$\ft_j \et_i \cdot u=C(u,q)f_j e_i\cdot u=C(u,q)e_i f_j\cdot u=0$,
with $C(u,q)\in\BQ(q)$.
Hence we have shown:
$$
\et_i \ft_j \cdot u=\ft_j \et_i \cdot u,
\qquad u\in B,\;\;i\not=j.
\eqno(5.14)
$$
\thm{6}. 
\point(i) The crystal graph of $V(m,n;a,b)$ is connected
if and only if $m\not=n$.
\point(ii) If $m=n$, it has two components.
\endthm 
\proof
Suppose $m\not=n$. Then by \reflem{5.3},
$V(m,n;a,b)$ has an escalator, and looking at \refeq{5.15,5.16}
we see that its crystal graph has an infinite subgraph ${\cal E}$
of the form
$$
\cdots\rightarrow b^p_{jk}
\rightarrow b^p_{j+1,k}\rightarrow b^{p-1}_{jk}
\rightarrow b^{p-1}_{j+1,k}\rightarrow\cdots
\eqno()
$$
or
$$
\cdots\rightarrow b^p_{jk}
\rightarrow b^p_{j,k+1}\rightarrow b^{p-1}_{jk}
\rightarrow b^{p-1}_{j,k+1}\rightarrow\cdots
\eqno()
$$
To prove the connectedness of the crystal graph, it is enough
to show that there exists a path from every vertex
$b^p_{j'k'}$ to ${\cal E}$. This last fact is proved as follows.
If $j>0$, $b^p_{j'k'}=(\ft_1)^{2j+k-m}b^p_{0,m-j}$, since
$b^p_{0,m-j}\in\Ker e_1\mod qL$, and similarly,
$b^p_{0,m-j}=(\ft_0)^{2n-m+j}b^{p+2n-m+j}_{mn}$. Therefore,
$\forall p\in\BZ$, the vertex $b^p_{j'k'}$ is connected
to the SE corner of degree $p+2n-m+j$, i.e. $b^{p+2n-m+j}_{mn}$.
On the other hand, because the escalator ${\cal E}$ is infinite,
and contains vertices of all degrees $p\in\BZ$,
there exists a vertex of ${\cal E}$ connected to the same SE corner
of degree $p+2n-m+j$. Thus, we have shown that $m\not=n$
is a sufficient condition for connectedness. We now prove that
it is also a necessary condition, by showing that, if $m=n$,
$\forall p\in\BZ$, $\forall j,k$, the vertices
$b^p_{jk}$ and $b^{p-1}_{jk}$ are not connected.
Assume for a contradiction that there exists a monomial
$P(\et_i,\ft_i)$, $i=0,1$, such that
$$
b^{p-1}_{jk}=P(\et_i,\ft_i)b^p_{jk}.
\eqno(5.17)
$$
Consider the length of the word $P$,
i.e. the number of its letters $\et_i$, $\ft_i$. Let $P_0$ be
the monomial of minimal length satisfying \refeq{5.17}. Then
$P_0$ is a function of the letters $\ft_i$ only, not the $\et_i$.
Indeed, suppose that $P_0$ involves both $\et_i$ and $\ft_i$.
Then by virtue of \refeq{5.14} above, one can permute the
factors $\ft_i$ and $\et_j$ for $i\not=j$, without changing the
length, until one of the sequences $\et_i\ft_i$, $\ft_i\et_i$,
$i=0,1$ appears. (Such a sequence is bound to appear, for the
only possibility to avoid this is when
$$
P_0=\et_0^{n_0}\ft_1^{n_1}\quad{\rm or}\quad\et_1^{m_1}\ft_0^{m_0}.
\eqno(5.18)
$$
But this is forbidden by \refeq{5.17}: the eigenvalue of
$t_1$ acting on $b^p_{jk}$ and $b^{p-1}_{jk}$ being the same,
we must have $t_1 P_0 t_1^{-1} = P_0$, while this does not hold
in \refeq{5.18}.)

Now $\et_i\ft_i=\ft_i\et_i=1$, and hence we get a monomial
satisfying \refeq{5.17} shorter than $P_0$, contradicting
minimality. Taking into account the fact that there is
a difference of 1 in the degrees $p$ and $p-1$ of the vectors
in \refeq{5.17}, so that we must also have $[d,P_0]=-P_0$,
which means that $P_0$ contains $\ft_0$ with multiplicity 1,
we are left with the two possibilities
$P_0=\ft_0\ft_1$, $\ft_1\ft_0$. But these are also ruled out
by \reflem{5.3}.

It remains to prove (ii). We do this by studying
${\cal G}_m$, the crystal graph of $V(m,m;a,b)$.
More precisely we will show that $\forall p\in\BZ$,
$\forall j,k$, the vertices $b^p_{jk}$ and $b^{p-2}_{jk}$
are connected. This implies the statement in the
theorem, 
as the projection of the crystal graph on the $p=0$ plane
is connected.

The case $m=1$ is obvious by looking at ${\cal G}_1$.
Assume it has been shown for ${\cal G}_m$.
To prove it for ${\cal G}_{m+1}$, it is enough to show that
every one of the vertices of
${\cal G}_{m+1}\setminus {\cal G}_m$ has
the required property. But this is clear from an
inspection of the graph.
\qed
\section{Eigenstates of the \mbXXZ\ Hamiltonian}

\subsection{Basic constructions.}
The XXZ Hamiltonian is formally defined as
$$
\HX{}=-{1\over2}
\sum_{k=-\infty}^\infty\left(\sigma_{k+1}^x\sigma_k^x+
\sigma_{k+1}^y\sigma_k^y+\Delta\sigma_{k+1}^z\sigma_k^z\right),
\eqno(6.1)
$$
where $\sigma_k^x$, etc., are Pauli matrices acting at site $k$, and
the four-fermion coupling constant $\Delta$ is related to the quantum
algebra parameter $q$ by $\Delta=(q+q^{-1})/2$.
Similarly, the generator of the (two-sided) corner transfer matrix
has the formal definition
$$
\HC{}=-{q\over 1-q^2}
\sum_{k=-\infty}^\infty k\left(\sigma_{k+1}^x\sigma_k^x+
\sigma_{k+1}^y\sigma_k^y+\Delta\sigma_{k+1}^z\sigma_k^z\right).
\eqno(6.2)
$$
It is shown in \refto{FM92,Dav93} that the left-hand half ($k>0$)
of the CTM may be identified with the derivation $d$ of $\Uq$,
acting on a standard level-1 highest weight module.
There are two such modules, $V(\Lambda_0)$ and $V(\Lambda_1)$, which
correspond to the two possible choices of boundary condition in the
anti-ferromagnetic regime.
The right-hand half ($k<0$) of the CTM acts similarly, but in a dual
module, since the eigenvalues of this half of $\HC{}$ are negated.
Let $T$ be a shift by one lattice unit.
Then the naive definitions of $\HX{}$ and $\HC{}$ make natural the
identification
$$
{q\over1-q^2}\HX{}=T\cdot \HC{}\cdot T^{-1} - \HC{}
=T\cdot d\cdot T^{-1} - d.
\eqno(6.3)
$$
Given this, the space of states involves the level-0 modules
$V(\Lambda_i)\otimes V^*(\Lambda_j)$, $(i,j=0,1)$, \ie, loop modules.

We turn to the necessary definitions for the vertex operators (VOs).
They are intertwiners of $\Uq$ modules.
The appropriate VOs for $\HX{}$ intertwine the level-$1$ standard
modules and the loop module $V(1;a)$. In the sequel we use the
notations of \refto{DFJMN92}, thus we denote the variable $\st$ by
$z$ and the module $V(1;a)$ by $V_{za}$.
There are two basic types.
\point(i) Type I:
$$
\eqalign{
\tilde{\Phi}_\lambda^{\mu V}(z)&:
V(\lambda)\longrightarrow
V(\mu)\otimes V_z, \cr
\tilde{\Phi}^\lambda_{\mu V}(z)&:
V(\mu)\otimes V_z\longrightarrow V(\lambda). \cr}
\eqno(6.11)
$$
\point(ii) Type II:
$$
\eqalign{
\tilde{\Phi}_\lambda^{V\mu}(z)&:
V(\lambda)\longrightarrow
V_z\otimes V(\mu), \cr
\tilde{\Phi}^\lambda_{V\mu}(z)&:
V_z\otimes V(\mu)\longrightarrow V(\lambda). \cr}
\eqno(6.12)
$$
The existence and the uniqueness of such vertex operators, once the
overall normalisation is fixed, is proved in \refto{DJO92} in a
general setting.
Strictly speaking the image is a completion of the appropriate space,
but we shall not worry about this.
The grading is preserved, with the derivation
acting on the tensor products as $d\otimes1+1\otimes d$.

The translation operator $T$ uses the type-I VOs, but as $\Up$
intertwiners (formally, set $z=1$).
Type-I VOs preserve the crystal base, a property which is
important in making the physical connection between translation and
the action of these VOs \refto{Dav93}.
Explicitly, a shift to the right by one lattice unit is the
composition of maps:
$$
T:
V(\lambda)\otimes V^*(\lambda^\prime)
\mapright{\Phi_{\lambda}^{\mu V}\!\otimes\id}
V(\mu)\otimes V\otimes V^*(\lambda^\prime)
\mapright{\id\otimes\Phi_{V\lambda^\prime}^{*\mu^\prime}}
V(\mu)\otimes V^*(\mu^\prime)
\eqno(6.14)
$$
Note that translation by one lattice site changes the boundary
condition.

\subsection{Particle states.}
In \refto{DFJMN92}, the vacuum state is identified as the canonical
element of $V(\Lambda_0)\otimes V^*(\Lambda_0)$ corresponding to the
identity map of $\Hom(V(\Lambda_0),V(\Lambda_0))$:
$$
W_0
=\sum_k u_k\otimes u^*_k.
\eqno(6.21)
$$
Here $\{u_k\}$ is a basis of $V(\Lambda_0)$, and $\{u^*_k\}$ the dual
basis of $V^*(\Lambda_0)$.
$W_0$ is a trivial one-dimensional $\Uq$ module, whose
action is given by the co-unit $\varepsilon$.

The creation operator $\varphi^*_\pm(z)$ is defined by the action
$$
\varphi^*_\pm(z)\cdot(u\otimes u^*) =
\tilde{\Phi}^\mu_{V\lambda}(z) (v_\pm\otimes u)\otimes u^*.
\eqno(6.22)
$$
Plainly,
$\tilde{\Phi}^\mu_{V\lambda}(z)\otimes\id$ is an intertwining operator
$$
\tilde{\Phi}^\mu_{V\lambda}(z)\otimes\id:
V_z \otimes V(\Lambda_i)\otimes V^*(\Lambda_0)\longrightarrow
V(\Lambda_{1-i})\otimes V^*(\Lambda_0).
\eqno(6.23)
$$
We already noted that the trivial module $W_0$ is embedded
canonically in $V(\Lambda_0)\otimes V^*(\Lambda_0)$.
Now remember that the intertwiner $\tilde{\Phi}^\mu_{V\lambda}(z)$
is really a formal Laurent expansion:
$\tilde{\Phi}^\mu_{V\lambda}(z)(v_\pm\otimes u)=
\sum_{n\in\BZ}\tilde{\Phi}^\mu_{V\lambda,n}(z^n v_\pm\otimes u)$.
Therefore, from \refeq{6.23} and the properties of the co-unit, we
see that the one-particle states $\varphi^*_\pm(z)\cdot W_0$ are the
image under $(\tilde{\Phi}^{\Lambda_1}_{V\Lambda_0}(z)\otimes\id)$ of
$V_z\otimes W_0$, since $\{z^n\otimes v_\pm\}$ is the basis of $V_z$.
$V_z$ is irreducible and the map
$\tilde{\Phi}^{\Lambda_1}_{V\Lambda_0}(z)\otimes\id$ is a module
homomorphism, so the one-particle states form a module which we
identify with $V_z$.

Now we iterate this process: the two-particle states are the image
under
$\tilde{\Phi}_{V\Lambda_1}^{\Lambda_0}(z)$ $\otimes\,\id$
of $V_z \otimes V_z \otimes W_0$ and may be identified with
the tensor product $V_z \otimes V_z$.
We know that such tensor products are infinitely reducible, with
quotients $V(1,1;s,1)$, $s\in\CX$.
Therefore the $2$-particle states are also infinitely reducible, and
similarly for general $N$.
The complex number $s$ corresponds to the relative momentum of the
particles via $s=\exp({\rm i}(p_1-p_2))$.
In the Lie algebra case, it is shown in \refto{CP86} that the
variables $a_i$ must be unimodular in order that the modules be
unitarizable, corresponding to real momenta.

\subsection{Crystal base.}
It was conjectured in \refto{DFJMN92} that the n-particle states,
which are a set of linear maps in $\Hom(V(\Lambda_i),V(\Lambda_j))$,
preserve the crystal lattice, even though the type-II VOs which are
employed in their definition do not enjoy this property.
In fact, the space of states ${\cal
V}_{\lambda,\lambda^\prime}$ is defined in section 7.1 of
\refto{DFJMN92} using this assumption and it is observed that
$W_0\in{\cal V}_{\lambda,\lambda}$. Apart from the technical
consideration, one expects that a reasonable physical theory should
have this property since the conventional approach to the calculation
of physical quantities such as correlation functions is that they are
the infinite limit of a low-temperature expansion --- here an expansion
in the variable $q$ about $q=0$.
The theory of the crystal base is precisely about how one might
make such expansions even though the underlying field, $\BQ(q)$,
contains elements with arbitrarily large negative power of $q$.

Here we shall obtain some further partial results, extending the
proof to $1$-particle states and presenting further evidence to
support the conjecture in general.
We recall some results from \refto{DFJMN92}.
\lem{6.0}.
Let $T_\lambda\in\End(V(\lambda))$ denote the linear map
$$
T_\lambda u_\nu = q^{(2\rho,\lambda-\nu)} u_\nu,\qquad
u_\nu\in V(\lambda)_\nu.
\eqno()
$$
Then the lower crystal lattice $L^*(\lambda)$ of $V^*(\lambda)$ is
characterised as
$$
L^*(\lambda)=\{u\in V^*(\lambda)\mid
\langle u,T_\lambda L(\lambda)\rangle\}\in A(q).
\eqno()
$$
\endlem
\proof
This result is proved in \refto{DFJMN92}, proposition 6.2, for the
upper crystal lattice.
For the lower crystal lattice one simply uses \refeq{5.5:b}.
\qed

This has an implication for the vacuum state.
If one chooses the basis $\{u_k\}$ in \refeq{6.21} to be also a crystal
base of $V(\Lambda_0)$, then the elements of the dual basis provide a
crystal base when multiplied by appropriate powers of $q$:
$b^*_k=q^{-(2\rho,\lambda-\nu_k)}u^*_k$.
So
$$
W_0=\sum_k q^{(2\rho,\lambda-\nu_k)} b_k\otimes b^*_k.
\eqno(6.45)
$$
One sees that the crystal base of $W_0$ is simply
$b_{\Lambda_0}\otimes b^*_{\Lambda_0}$.
To treat the one-particle states $\varphi_\pm(z)\cdot W_0$,
we need a  preliminary idea.
\lem{6.1}.
Let $L(\lambda)$ be the lower crystal lattice of $V(\lambda)$, and
let $\tilde{\Phi}_{V\lambda}^{\mu}(z):
V_z \otimes V(\lambda)\longrightarrow V(\mu)$
be a type-II vertex operator, with respect to the lower co-product.
Then for any $u\in T_\lambda L(\lambda)$, we have
$\tilde{\Phi}_{V\lambda}^{\mu}(z)\cdot(v_\pm\otimes u)\in L(\mu)$.
\endlem
\proof
Without loss of generality we assume that $u$ is a weight vector of
weight $\nu$.
We use induction on the height $(2\rho,\lambda-\nu)$, noting
that  $(2\rho,\lambda-\nu)=m_0+m_1$
when $\lambda-\nu=m_0\alpha_0+m_1\alpha_1$.
We need only consider the case of
$\tilde{\Phi}_{V\Lambda_0}^{\Lambda_1}(z)$;
the intertwiner $\tilde{\Phi}_{V\Lambda_1}^{\Lambda_0}(z)$
may be obtained from it by a Dynkin diagram automorphism.
The case that the height is zero involves only the highest weight
vector $u_{\Lambda_0}$.
Consider the images
$e_i\cdot\tilde{\Phi}_{V\Lambda_0}^{\Lambda_1}(z)
(v_\pm\otimes u_{\Lambda_0})=
\tilde{\Phi}_{V\Lambda_0}^{\Lambda_1}(z)
(e_i\cdot v_\pm\otimes u_{\Lambda_0})$ by the
intertwining property.
Substituting for $e_i\cdot v_\pm$ in each case, we obtain the equations
$$
\eqalignno{
e_0\cdot\tilde{\Phi}_{V\Lambda_0}^{\Lambda_1}(z)
(v_{+}\otimes u_{\Lambda_0})
&=z\tilde{\Phi}_{V\Lambda_0}^{\Lambda_1}(z)
(v_{-}\otimes u_{\Lambda_0}), &(6.31:a)\cr
e_0\cdot\tilde{\Phi}_{V\Lambda_0}^{\Lambda_1}(z)
(v_{-}\otimes u_{\Lambda_0})
&=0,  &(6.31:b)\cr
e_1\cdot\tilde{\Phi}_{V\Lambda_0}^{\Lambda_1}(z)
(v_{+}\otimes u_{\Lambda_0})
&=0,  &(6.31:c)\cr
e_1\cdot\tilde{\Phi}_{V\Lambda_0}^{\Lambda_1}(z)
(v_{-}\otimes u_{\Lambda_0})
&=\tilde{\Phi}_{V\Lambda_0}^{\Lambda_1}(z)
(v_{+}\otimes u_{\Lambda_0}),  &(6.31:d)\cr}
$$
which implies that each weight component of the image is in a
spin one-half $\U$ sub-module with respect to both $i=0$ and $i=1$.
Therefore the modified Chevalley generators $\ft_i$ have the same
action as the $f_i$ on the image, which proves the initial step of the
induction by \refeq{5.5:a}.

For the inductive step, consider an arbitrary weight vector\break
$u\in q^{(2\rho,\lambda-\nu)}L(\Lambda_0)_\nu$, whose height is one
more than the current level of the induction.
Then $u=\ft_i u^\prime$ for some $i$ and some
$u^\prime$, and the latter has the decomposition
$u^\prime=\sum_{j\ge0}f_i^{(j)}u_j$, $u_j\in\Ker(e_i)~\forall~j$.
So $u=\sum_{j\ge0}f_i^{(j+1)}u_j$, and it is sufficient to consider
the case that there is only one term in the sum, \ie,
$u^\prime=f_i^{(j)}u_j$, so that $f_i u^\prime=[j+1]u$.
The crucial point is that
$$
q^{-1}u^\prime\in
q^{(2\rho,\lambda-\nu^\prime)}L(\Lambda_0)_{\nu^\prime}.
\eqno(6.41)
$$
{}From the intertwining property
$f_i\cdot\tilde{\Phi}_{V\Lambda_0}^{\Lambda_1}
(z)(v_{\pm}\otimes u^\prime)
=\tilde{\Phi}_{V\Lambda_0}^{\Lambda_1}(z)
(f_i\cdot v_{\pm}\otimes u^\prime)
+\tilde{\Phi}_{V\Lambda_0}^{\Lambda_1}(z)
(t_i\cdot v_{\pm}\otimes f_i\cdot u^\prime)$,
which implies
$$
\tilde{\Phi}_{V\Lambda_0}^{\Lambda_1}(z)
(v_{\pm}\otimes u) =
{f_i\cdot\tilde{\Phi}_{V\Lambda_0}^{\Lambda_1}
(z)(v_{\pm}\otimes u^\prime)
-\tilde{\Phi}_{V\Lambda_0}^{\Lambda_1}(z)
(f_i\cdot v_{\pm}\otimes u^\prime) \over q^{\pm1}[j+1]}.
\eqno(6.42)
$$
The induction hypothesis applies to the right hand side.
Moreover, $e_i^{j+1}u^\prime=0$ and either $e_iv_{\pm}=0$ or
$e_i^2v_{\pm}=0$. By the intertwining property,
$$
e_i^M\cdot\tilde{\Phi}_{V\Lambda_0}^{\Lambda_1}(z)
(v_{\pm}\otimes u^\prime)=0,  \qquad
\left\{\eqalign{
M&=j+1 \quad {\rm if} \quad e_iv_{\pm}=0, \cr
M&=j+2 \quad {\rm if} \quad e_i^2v_{\pm}=0. \cr}\right.
\eqno(6.43)
$$
Write $\tilde{\Phi}_{V\Lambda_0}^{\Lambda_1}(z)(v_{\pm}\otimes u^\prime)
=\sum_n w_n$.
It is easy to see that
$(f_i/[j+1])\cdot w_n\approx q^{j-M+1} \ft_i\cdot w_n$, giving
$$
(f_i/[j+1])\cdot\tilde{\Phi}_{V\Lambda_0}^{\Lambda_1}
(z)(v_{\pm}\otimes u^\prime)
\approx
q^{j+1-M}\ft_i\cdot\tilde{\Phi}_{V\Lambda_0}^{\Lambda_1}
(z)(v_{\pm}\otimes u^\prime).
\eqno(6.44)
$$
Similar considerations apply to the other term on the right hand side
of \refeq{6.41}.
The result follows immediately.
\qed
{}From \refeq{6.45} and the preceeding lemma, it follows trivially that
\thm{6.1}.
The one-particle states $\varphi_\pm(z)\cdot
W_0 \in\Hom(V(\Lambda_0),V(\Lambda_1))$ preserve the crystal lattice.
\endthm

One sees from the foregoing the difficulty in proceeding further.
Formally any state is a sum of the form
$$
\Psi(z_1,\cdots,z_n)=\sum_{k,l}
c_{k,l}(z_1,\cdots,z_n) b_k\otimes b_l^*
\eqno(6.50)
$$
and we want to show that $c_{k,l}(z_1,\cdots,z_n)\in A(q)$.
These coefficients are obtained from the
definition \refeq{6.22} as infinite sums, with individual terms
involving arbitrary negative powers of $q$.
We expect that they should be analytic functions in some
annulus which contains the physical region $|z_1|=\cdots=|z_n|=1$.
But, as discussed in \refto{DFJMN92}, the formal definition of the
necessary matrix elements is by meromorphic continuation from the
region $|z_1|\gg\cdots\gg|z_n|$, around poles at
$z_i/z_{i+1}=q^{-2}$.
Some method to perform the continuation is required for further
progress to be made.
Probably the necessary information is contined in the fact
that the creation-annihilation operators provide a lattice
realization of the Zamolodchikov algebra.
For example, the creation operators are known to satisfy (eqn.(7.10a)
of \refto{DFJMN92})
$$
\varphi^*_{\epsilon_1}(z_1)\varphi^*_{\epsilon_2}(z_2)=
\sum_{\epsilon_1^\prime\epsilon_2^\prime}
\big(R_{V^*V^*}(z_1/z_2)\big)_{\epsilon_1^\prime\epsilon_2^\prime}
^{\epsilon_1\epsilon_2}\
\varphi^*_{\epsilon_1^\prime}(z_2)
\varphi^*_{\epsilon_2^\prime}(z_1)
\eqno()
$$
where the $R$ matrix (eqn.(6.18 ) of \refto{DFJMN92}) is meromorphic
in the annulus $q^2<|z|<q^{-2}$.
Moreover, the information about analyticity comes from general
considerations and is needed to compute the normalision of the $R$
matrix --- which contains most of the analytic information --- by the
method of Frenkel and Reshetikhin \refto{FR92}.
The commutation relations in turn place severe constraints on the
coefficients appearing in \refeq{6.50}, possibly sufficiently strong
to allow a proof of the conjecture about preservation of the crystal
base for arbitrary $n$.
But at present we are unable to find a way to construct such a proof.
\section{Conclusions}

This paper has been concerned mainly with the study of loop modules
of the quantum algebra $\Uq$.
The motivation for this study is the importance of representation
theory in the solution of two-dimensional lattice models of
statistical mechanics.
The simplest loop modules --- the affinization of an irreducible $\U$
module --- are of central importance in the papers cited in the
introduction and are extensively studied in \refto{KKMMNN92}.
Nevertheless, some important and interesting questions are raised even
in the solution of the six-vertex model.
In \refto{DFJMN92} one sees that new features emerge in the
$n$-particle sector with $n\ge2$.
The $n$-particle sectors are built on a certain
symmetrisation, under the $R$-matrix symmetry, of the tensor
product of $n$ copies of a simple $\Uq$ module $V_z$.
Unlike the 1-particle sector, an understanding of the structure
of these sectors at $q=0$ left some unexplained puzzles.

In the present paper we have defined rather general quantum loop
modules, although certainly not the most general possible.
In this regard our choice is dictated by the physically interesting
questions raised by the cited works.
Since the $\Uq$ modules constructed herein are also $\Up$ modules, it
is natural that our treatment should parallel the works of Chari and
Pressley to some extent.
In fact, the conditions we find for the modules to be cyclic are
exactly those of the corresponding Lie algebra case; similarly the
conditions for irreducibility are the same as for $\Up$ modules.
One sees however that there is a sharp distinction
between the irreducibility properties of loop modules constructed by
affinizing a tensor product of irreducible $\U$ modules
$V_1,\cdots,V_n$, and the tensor product of the affinizations of the
individual $V_j$.
The latter is highly reducible.

The crystal base theory of $\Uq$ modules has great significance in
the physical interpretation of the theory.
It is the connecting link between the use of representation theory
and the traditional method of low-temperature expansions.
For this reason we have studied the next simplest case
after $V(m;a)$ in some detail.
In particular, the modules $V(m,n;a,b)$ do not have a crystal base in
general unless $m=n$. In the latter case, the crystal graph is
also not connected, although the module is irreducible.
Presumably these properties extend to general loop modules
$V(\bl;\ba)$. This would be entirely consistent with the findings of
Reshetikhin \refto{R91} (using the Bethe Ansatz) and Idzumi \etal
\refto{IIJMNT92} (using the quantum affine symmetry) for the higher
spin generalization of the six-vertex model.
The important point is that even though the basic representation are
all spin $k/2$, $k\ge2$, the elementary excitations of the system are
all spin $1/2$ particles.

Section 6 also contains incomplete results.
Nevertheless considerable new light is thrown onto the solution of
the six-vertex model using the quantum affine symmetry, presented in
\refto{DFJMN92}.
Here we just recall our results for the two-particle sector.
We saw in section 6 that it is isomorphic to a tensor
product $V_z \otimes V_z$, which is highly
reducible into copies of $V(1,1;s,1)$, $s\in\CX$.
Physically, this means that
states of given total momentum $s=s_1s_2$ --- recall that
$s_j=\exp(iu_j)$, $u\in\BR$ --- still are infinitely degenerate, with
internal momentum given by the ratio $s=s_1/s_2$.
This fact is exactly mirrored in the possibilities for constructing
the crystal base.
It is well-known that the tensor product of irreducible
finite-dimensional $\Up$ modules is irreducible.
But the crystal base for the $n$-particle sectors which was proposed
in \refto{DFJMN92} has an infinite number of disconnected components,
which correspond to only the fraction $1/n!$ of the number of
such components which might be expected from taking the tensor product
of $n$ copies of the crystal base for $V_z$.
There is no contradiction here: it simply illustrates the completely
different reducibility properties of the tensor product of loop
modules.

\section*{Acknowledgements}

This research was partly supported by ARC grant A69130579/92.
D.A. would like to thank the Mathematics Department of the ANU for
their warm hospitality.

\section*{References}

\def\SMD{Sov. Math. Dokl.}

\refis{ABF84} \refjl
{G. E. Andrews, R. J. Baxter and P. J. Forrester}
{\JSP}{35}{(1984), 193}

\refis{Bax76} \refjl
{R. J. Baxter}
{\JSP}{15}{(1976), 485}

\refis{Bax77} \refjl
{R. J. Baxter}
{\JSP}{17}{(1977), 1}

\refis{BPZ84} \refjl
{A. A. Belavin, A. M. Polyakov and A. B. Zamolodchikov}
{\NPB}{241}{(1984) 333}

\refis{Ch86} \refjl
{V. Chari}
{\IM}{85}{(1986) 317}

\refis{CP86} \refjl
{V. Chari and A. Presley}
{\MA}{275}{(1986) 87}

\refis{CP87} \refjl
{V. Chari and A. Presley}
{\MA}{277}{(1987) 543}

\refis{CP91} \refjl
{V. Chari and A. Presley}
{\CMP}{142}{(1991) 261}

\refis{Dav93} \refjl
{B. Davies}{\JPA}{}{(1993), in press}

\refis{DFJMN92} \refjl
{B. Davies, O. Foda, M. Jimbo, T. Miwa and A. Nakayashiki}
{\CMP}{151}{(1993), 89}

\refis{DJKMNO87} \refjl
{E. Date, M. Jimbo, A. Kuniba, T. Miwa and M. Okado}
{\NPB}{290}{(1987), 231}

\refis{DJKMNO89} \refjl
{E. Date, M. Jimbo, A. Kuniba, T. Miwa and M. Okado}
{\LMP}{17}{(1987), 69}

\refis{DJO92} \refjl
{E. Date, M. Jimbo and M. Okado}
{Osaka Univ. Math. Sci. Preprint}{1}{(1992)}

\refis{DO93} \refjl
{E. Date and M. Okado}
{Osaka Univ. Math. Sci. Preprint}{1}{(1993)}

\refis{Dr87} \refbk
{V.G. Drinfeld}
{Quantum groups, Proc. ICM Berkeley}{(1987), 798}

\refis{Dr88} \refjl
{V. G. Drinfeld}{\SMD}{36}{(1988) 212}

\refis{FC91} \refjl
{G. Felder and A. LeClair}
{\RIMS}{799}{(1991)}

\refis{FJMMN93} \refjl
{O. Foda, M. Jimbo, K. Miki, T. Miwa and A. Nakayashiki}
{\RIMS}{}{}

\refis{FM92} \refbk
{O. Foda and T. Miwa}
{Infinite Analysis}
{World Scientific (1992)}

\refis{FR92} \refjl
{I. B. Frenkel and N. Yu. Reshetikhin}
{\CMP}{149}{(1992), 1}

\refis{IIJMNT92} \refjl
{M. Idzumi, K. Iohara, M. Jimbo, T. Miwa, T. Nakayashima and
T. Tokihiro}
{\IJMPA}{}{(1992), in press}

\refis{J85} \refjl
{M. Jimbo}{\LMP}{10}{(1985), 63}

\refis{J86} \refjl
{M. Jimbo}{\LMP}{11}{(1986), 247}

\refis{J89} \refjl
{M. Jimbo}{\IJMPA}{4}{(1989), 3759}

\refis{J92} \refbk
{M. Jimbo}
{Nankai Lecture Notes on Mathematical Physics, ed. Ge Mo-Lin}
{World Scientific (1992)}

\refis{JMMN92} \refjl
{M. Jimbo, K. Miki, T. Miwa and A. Nakayashiki}
{\PLA}{168}{(1992) 256}

\refis{JMN92} \refjl
{M. Jimbo, T. Miwa and A. Nakayashiki}
{\JPA}{}{(1992), in press}

\refis{JMO92} \refjl
{M. Jimbo, T. Miwa and Y. Ohta}
{\IJMPA}{}{(1992), in press}

\refis{Ka90} \refjl
{M. Kashiwara}
{\CMP}{133}{(1990), 249}

\refis{KKMMNN92} \refbk
{S.J. Kang, M. Kashiwara, K.C. Misra, T. Miwa, T. Nakashima and
A. Nak\-aya\-sh\-iki}{Infinite Analysis}{World Scientific (1992)}

\refis{Ma79} \refbk
{I. G. MacDonald}
{Symmetric functions and Hall polynomials}
{Oxford: Clarendon Press (1979)}

\refis{R91} \refjl
{N. Yu. Reshetikhin}
{\JPA}{24}{(1991), 3299}

\listreferences

\end{document}

\bye